\documentclass[10pt,journal,compsoc]{IEEEtran}

\usepackage{todonotes}
\usepackage{soul}
\usepackage{mathtools}
\usepackage{amsmath}
\usepackage{multirow}
\usepackage{booktabs}
\usepackage{subfig}
\usepackage{comment}

\usepackage{array}
\usepackage{rotating}
\usepackage{multirow}
\usepackage{xspace}
\usepackage{float}

\makeatletter

\providecommand{\tabularnewline}{\\}

\makeatother

\usepackage{amsmath}
\usepackage{amssymb}
\usepackage{amsfonts}
\usepackage{tikz}
\def\checkmark{\tikz\fill[scale=0.3](0,.35) -- (.25,0) -- (1,.7) -- (.25,.15) -- cycle;} 

\ifCLASSOPTIONcompsoc
  \usepackage[nocompress]{cite}
\else
  \usepackage{cite}
\fi

\ifCLASSINFOpdf
  \graphicspath{{./Figures/}{../jpeg/}}
  \DeclareGraphicsExtensions{.pdf,.jpeg,.png}
\else
\fi

\newcommand{\etal}{et al.\@\xspace} 

\newcommand{\ParLabel}[1]{\vspace{\topsep}\noindent{\sf\bfseries\small #1:}}

\begin{document}
%
\title{A Survey of Calibration Methods for Optical See-Through Head-Mounted Displays}
%
%
%
%

\author{Jens Grubert (\IEEEmembership{Member,~IEEE}),
        Yuta Itoh (\IEEEmembership{Member,~IEEE}),
        Kenneth Moser, \\
        and~J. Edward Swan II (\IEEEmembership{Senior Member,~IEEE})
        \IEEEcompsocitemizethanks{\IEEEcompsocthanksitem J. Grubert is with the Department of Electrical Engineering and Computer Science, Coburg University, Germany.\protect\\
E-mail: jg@jensgrubert.de
\IEEEcompsocthanksitem Yuta Itoh is with the Department of Information and Computer Science, Keio University, Japan.\protect\\
E-mail: itoh@imlab.ics.keio.ac.jp
\IEEEcompsocthanksitem Kenneth Moser is in Research and Development at Marxent Labs LLC, USA.\protect\\
E-mail: moserk@acm.org
\IEEEcompsocthanksitem J. Edward Swan II is with the Center for Advanced Vehicular Systems, Mississippi State University, USA. \protect\\
E-mail: : swan@acm.org}
\thanks{}
\thanks{}}

%
%

\markboth{author version}%
{Shell \MakeLowercase{\textit{et al.}}: Bare Advanced Demo of IEEEtran.cls for Journals}
%



\IEEEtitleabstractindextext{%
\begin{abstract}
Optical see-through head-mounted displays (OST HMDs) are a major output medium for Augmented Reality, which have seen significant growth in popularity and usage among the general public due to the growing release of consumer-oriented models, such as the Microsoft Hololens.  Unlike Virtual Reality headsets, OST HMDs inherently support the addition of computer-generated graphics directly into the light path between a user's eyes and their view of the physical world.  As with most Augmented and Virtual Reality systems, the physical position of an OST HMD is typically determined by an external or embedded 6-Degree-of-Freedom tracking system.  However, in order to properly render virtual objects, which are perceived as spatially aligned with the physical environment, it is also necessary to accurately measure the position of the user's eyes within the tracking system's coordinate frame.  For over 20 years, researchers have proposed various calibration methods to determine this needed eye position.  However, to date, there has not been a comprehensive overview of these procedures and their requirements.  Hence, this paper surveys the field of calibration methods for OST HMDs.  Specifically, it provides insights into the fundamentals of calibration techniques, and presents an overview of both manual and automatic approaches, as well as evaluation methods and metrics.  Finally, it also identifies opportunities for future research.
\end{abstract}

\begin{IEEEkeywords}
augmented reality, head-mounted displays, optical see-through calibration
\end{IEEEkeywords}}

\maketitle

\global\long\def\argmax{\operatornamewithlimits{argmax}}
 \global\long\def\argmin{\operatornamewithlimits{argmin}}

\global\long\def\Matrix#1{\mathtt{#1}}

\global\long\def\Rotation{\Matrix R}

{\small{}{}}\global\long\def\Transpose{\mathrm{T}}

\global\long\def\Translation{\mathbf{t}}

\global\long\def\Projection{\Matrix P}

\global\long\def\Identity{\Matrix I}

\global\long\def\screen{\mathbf{s}}

\global\long\def\point{\mathbf{p}}

\global\long\def\Intrinsic{\Matrix K}

\global\long\def\Screen{\mathbf{S}}

\global\long\def\Real{\mathbb{R}}

\global\long\def\OperatorA#1#2{{#1_{{\scriptscriptstyle #2}}}}

\global\long\def\OperatorAB#1#2#3{{_{{\scriptscriptstyle #3}}^{{\scriptscriptstyle #2}}#1}}

\global\long\def\RotationAB#1#2{\OperatorAB{\Rotation}{#1}{#2}}

\global\long\def\TransAB#1#2{\OperatorAB{\Translation}{#1}{#2}}

\global\long\def\ProjectionAB#1#2{\OperatorAB{\Projection}{#1}{#2}}

\global\long\def\PointA#1{\OperatorA{\point}{#1}}

\global\long\def\IntrinsicA#1{\OperatorAB{\Intrinsic}{#1}{}}

\global\long\def\ScreenA#1{\OperatorA{\Screen}{#1}}

\global\long\def\Homogenuos#1{\widetilde{#1}}

\global\long\def\eyeCoordinate{E}

\global\long\def\cameraCoordinate{T}

\global\long\def\worldCoordinate{H}

\global\long\def\screenCoordinate{S}

\global\long\def\spaamCoordinate{\eyeCoordinate_{0}}

\global\long\def\IntrinsicSPAAM{\IntrinsicA{\spaamCoordinate}}

\global\long\def\IntrinsicEye{\IntrinsicA{\eyeCoordinate}}

\global\long\def\ProjectionEye{\OperatorA{\Projection}{\eyeCoordinate}}


\global\long\def\ScreenEye{\ScreenA{\eyeCoordinate}}

\global\long\def\ProjectionSPAAMtoEye{\ProjectionAB{\spaamCoordinate}{\eyeCoordinate}}

\global\long\def\ProjectionEyeToSPAAM{\ProjectionAB{\eyeCoordinate}{\spaamCoordinate}}

\global\long\def\ProjectionWorldToEye{\ProjectionAB{\worldCoordinate}{\eyeCoordinate}}

\global\long\def\pixelScaling{\alpha}

\global\long\def\eyeCenter{\mathbf{o}_{\eyeCoordinate}}

\global\long\def\screenSPAAMCenter{\TransAB{\screenCoordinate}{\spaamCoordinate}}

\global\long\def\screenEyeCenter{\TransAB{\screenCoordinate}{\eyeCoordinate}}

\global\long\def\screenEyeCenterX{x}

\global\long\def\screenEyeCenterY{y}

\global\long\def\screenEyeCenterZ{z}

\global\long\def\pointSPAAM{\PointA{\spaamCoordinate}}

\global\long\def\pointEye{\PointA{\eyeCoordinate}}

\global\long\def\pointWorld{\PointA{\worldCoordinate}}

\global\long\def\RotationFromWorldToCamera{\RotationAB{\worldCoordinate}{\cameraCoordinate}}

\global\long\def\RotationFromWorldToEye{\RotationAB{\worldCoordinate}{\eyeCoordinate}}

\global\long\def\RotationFromWorldToSPAAMss{\RotationAB{\worldCoordinate}{\spaamCoordinate}}

\global\long\def\RotationFromWorldToScreen{\RotationAB{\worldCoordinate}{\screenCoordinate}}

\global\long\def\RotationFromCameraToEye{\RotationAB{\cameraCoordinate}{\eyeCoordinate}}


\global\long\def\TransFromWorldToSPAAMs{\TransAB{\worldCoordinate}{\spaamCoordinate}}

\global\long\def\TransFromWorldToCamera{\TransAB{\worldCoordinate}{\cameraCoordinate}}

\global\long\def\TransFromWorldToEye{\TransAB{\worldCoordinate}{\eyeCoordinate}}

\global\long\def\TransFromEyeToWorld{\TransAB{\eyeCoordinate}{\worldCoordinate}}

\global\long\def\TransFromWorldToScreen{\TransAB{\worldCoordinate}{\screenCoordinate}}

\global\long\def\TransFromCameraToEye{\TransAB{\eyeCoordinate}{\cameraCoordinate}}


\global\long\def\TransFromSPAAMToEye{\TransAB{\spaamCoordinate}{\eyeCoordinate}}

\global\long\def\TransFromScreenToEyeOld{\TransAB{\screenCoordinate}{\spaamCoordinate}}

\global\long\def\TransFromScreenToEye{\TransAB{\screenCoordinate}{\eyeCoordinate}}

\global\long\def\TransFromEyeToSPAAMs{\TransAB{\eyeCoordinate}{\spaamCoordinate}}

\global\long\def\TransFromEyeToSPAAMsX{\Delta x}

\global\long\def\TransFromEyeToSPAAMsY{\Delta y}

\global\long\def\TransFromEyeToSPAAMsZ{\Delta z}

\global\long\def\PixelCvtMatrix{\mathbf{A}}

\global\long\def\Limbus{L}
 \global\long\def\radius{r}

\global\long\def\LimbusRadius{\OperatorA{\radius}{\Limbus}}
 \global\long\def\EyeRadius{\OperatorA{\radius}{\eyeCoordinate}}

\global\long\def\LimbusEllipse{\mathbf{Q}}

\global\long\def\NormalizedEllipse{\mathbf{Q}_{e}}

\global\long\def\EigenValMat{\boldsymbol{\Lambda}}
 \global\long\def\EigenVecMat{\mathbf{U}}

\global\long\def\EigenVec{\mathbf{e}}
 \global\long\def\EigenVal{\lambda}
 \global\long\def\Sign{s}

\global\long\def\EigenVecOne{\EigenVec_{1}}
 \global\long\def\EigenValOne{\alpha}
 \global\long\def\SignOne{\Sign_{1}}

\global\long\def\EigenVecTwo{\EigenVec_{2}}
 \global\long\def\EigenValTwo{\beta}
 \global\long\def\SignTwo{\Sign_{2}}

\global\long\def\EigenVecThree{\EigenVec_{3}}
 \global\long\def\EigenValThree{\gamma}
 \global\long\def\SignThree{\Sign_{3}}

\global\long\def\LimbusPosition{\TransAB{\Limbus}{\cameraCoordinate}}
 \global\long\def\GazeNormal{\OperatorA{\mathbf{n}}{\cameraCoordinate}}

\global\long\def\Edge{E}

\global\long\def\Image{\mathbf{I}}

\global\long\def\EdgeImage{\OperatorA{\Image}{\Edge}}

\global\long\def\Score{S}

\global\long\def\RANSACEllipse{\mathbf{Q}_{\mathrm{k}}}

\global\long\def\RANSACScore{\Score_{k}}

\global\long\def\BestEllipse{\OperatorA{\mathbf{Q}}{\mathrm{\mathrm{local\_best}}}}

\global\long\def\BestScore{\OperatorA{\Score}{\mathrm{local\_best}}}

\global\long\def\GlobalyBestEllipse{\OperatorA{\mathbf{Q}}{\mathrm{\mathrm{\mathrm{global\_best}}}}}

\global\long\def\GlobalyBestScore{\OperatorA{\Score}{\mathrm{global\_best}}}

\global\long\def\CameraCenter{c}

\global\long\def\Width{w}

\global\long\def\Height{h}

\global\long\def\CalibCameraCoordinate{C}

\global\long\def\CameraIndex{k}

\global\long\def\CameraIndexTwo{j}

\global\long\def\PointIndex{i}

\global\long\def\KthCalibCameraCoordinate{\CalibCameraCoordinate_{\CameraIndex}}

\global\long\def\JthCalibCameraCoordinate{\CalibCameraCoordinate_{\CameraIndexTwo}}

\global\long\def\TransFromScreenToWorld{\TransAB{\screenCoordinate}{\worldCoordinate}}

\global\long\def\TransFromScreenToCameraPrime{\TransAB{\screenCoordinate}{\CalibCameraCoordinate_{\CameraIndexTwo}}}

\global\long\def\Num{N}

\global\long\def\CameraNum{\OperatorA{\Num}{\CalibCameraCoordinate}}

\global\long\def\PointNum{\OperatorA{\Num}P}

\global\long\def\Scale{s}

\global\long\def\Mean#1{\overline{#1}}

\global\long\def\TransFromScreenToWorldMean{\Mean{\TransFromScreenToWorld}}

\global\long\def\TransFromScreenToWorldPerPointMean{\Mean{\TransFromScreenToWorld^{\PointIndex}}}

\global\long\def\RotationFromScreenToCameraPerCamera{\RotationAB{\screenCoordinate}{\KthCalibCameraCoordinate}}

\global\long\def\TransFromScreenToWorldPerCameraPerPoint{\TransFromScreenToWorld^{\PointIndex\CameraIndex}}

\global\long\def\TransFromScreenToCameraPerCamera{\TransAB{\screenCoordinate}{\KthCalibCameraCoordinate}}

\global\long\def\TransFromScreenToCameraPerCameraPerPoint{\TransFromScreenToCameraPerCamera^{\PointIndex}}

\global\long\def\TransFromScreenToWorldPerCamera{\TransFromScreenToWorld^{\CameraIndex}}

\global\long\def\TransFromScreenToCameraPerCameraPrime{\TransFromScreenToCameraPerCamera}

\global\long\def\TransFromCameraToWorldPerCamera{\TransAB{\KthCalibCameraCoordinate}{\worldCoordinate}}

\global\long\def\TransFromScreenToWorldPerCameraPerPointPrime{\TransFromScreenToWorld^{\PointIndex\CameraIndexTwo}}

\global\long\def\TransFromScreenToCameraPerCameraPerPointPrime{\TransFromScreenToCameraPerCamera^{\PointIndex\CameraIndexTwo}}

\global\long\def\TransFromScreenToWorldPerCameraPrime{\TransFromScreenToWorld^{\CameraIndexTwo}}

\global\long\def\RotationFromCameraToWorldPerCamera{\RotationAB{\KthCalibCameraCoordinate}{\worldCoordinate}}

\global\long\def\RotationFromCameraToWorldPerCameraPrime{\RotationAB{\JthCalibCameraCoordinate}{\worldCoordinate}}

\global\long\def\TransFromCameraToWorldPerCameraPrime{\TransAB{\JthCalibCameraCoordinate}{\worldCoordinate}}

\global\long\def\TransFromScreenToCameraPerCameraPrime{\TransFromScreenToCameraPrime}

\global\long\def\TmpValOne{\mathbf{a}^{\PointIndex\CameraIndex}}
 \global\long\def\TmpValTwo{\mathbf{b}^{\PointIndex\CameraIndex}}

\global\long\def\TmpValOnePerCamera{\mathbf{a}^{\CameraIndex}}
 \global\long\def\TmpValTwoPerCamera{\mathbf{b}^{\CameraIndex}}

\global\long\def\MarkerCoordinate{M}

\global\long\def\Estimate#1{\widehat{#1}}

\global\long\def\ScaleLinearEstimate{\Estimate{\Scale}}

\global\long\def\DcmToQ#1{\textrm{dcm2q}\left(#1\right)}

\global\long\def\QToDcm#1{\textrm{q2dcm}\left(#1\right)}

\global\long\def\ImagePoint{\mathbf{u}}
 \global\long\def\ThreeDPoint{\mathbf{x}}
 \global\long\def\DistortionCoeff{\boldsymbol{\theta}}
 \global\long\def\CameraPointPerPoint{\OperatorA{\ImagePoint^{i}}{\KthCalibCameraCoordinate}}
 \global\long\def\Distort#1#2{\mathrm{distort}(#1,#2)}

\global\long\def\ScreenPoint{\OperatorA{\ImagePoint^{i}}{\screenCoordinate}}
 \global\long\def\ScreenPointUndist{\OperatorA{\ImagePoint^{\star i}}{\screenCoordinate}}
 \global\long\def\CameraPoint{\OperatorA{\ImagePoint}{\CalibCameraCoordinate}}
 \global\long\def\CameraPoint{\mathrm{project}}

\global\long\def\CameraPointPerCamera{\OperatorA{\ImagePoint}{\KthCalibCameraCoordinate}}
 \global\long\def\ScreenPointPerCamera{\OperatorA{\ThreeDPoint}{\KthCalibCameraCoordinate}}

\global\long\def\RotationFromScreenToWorld{\RotationAB{\screenCoordinate}{\worldCoordinate}}

\global\long\def\CameraIndexSetPerPixel{C_{\PointIndex}}

\global\long\def\Quaternion{\mathbf{q}}

\global\long\def\ProjectionSlice{\mathbf{p}}

\global\long\def\ProjectionFunctionChar{\mathbf{f}}

\global\long\def\VarWithError#1{\hat{#1}}

\global\long\def\ProjectionFunctionWithErrorChar{\VarWithError{\ProjectionFunctionChar}}

\global\long\def\ProjectionFunction#1{\ProjectionFunctionChar\left(#1\right)}

\global\long\def\ProjectionFunctionWithError#1{\ProjectionFunctionWithErrorChar\left(#1\right)}

\global\long\def\ThreeDpointInE{\OperatorA{\mathbf{x}}{\eyeCoordinate}}

\global\long\def\ThreeDpointInW{\OperatorA{\mathbf{x}}{\worldCoordinate}}

\global\long\def\ThreeDpointInS{\OperatorA{\mathbf{x}}{\screenCoordinate}}

\global\long\def\ThreeDpointInWH{\Homogenuos{\ThreeDpointInW}}

\global\long\def\PixelPoint{\mathbf{u}}

\global\long\def\PixelPointByWorldPoint{\mathbf{u}_{\ThreeDpointInW}}

\global\long\def\PixelPointHomogeneous{\Homogenuos{\mathbf{u}}}

\global\long\def\PixelPointWithError{\VarWithError{\PixelPoint}}

\global\long\def\Length{\delta}
 \global\long\def\AngleOne{\theta}
 \global\long\def\AngleTwo{\varphi}

\global\long\def\LengthMax{L}
 \global\long\def\AngleMaxOne{\Theta}
 \global\long\def\AngleMaxTwo{\Phi}

\global\long\def\Distance{d}

\global\long\def\Error{\mathbf{e}}

\global\long\def\DistanceFunction#1#2{D\left(#1,#2\right)}

\global\long\def\Normalizer{V}

\global\long\def\CalibrationErrorChar{\mathrm{RMSE}}

\global\long\def\CalibrationErrorFunction#1{\CalibrationErrorChar\left(#1\right)}

\global\long\def\OriginalValue#1{#1^{\star}}

\global\long\def\ParametersCharSimple{\lambda}

\global\long\def\ParametersChar{\boldsymbol{\ParametersCharSimple}}

\global\long\def\Parameters{\OriginalValue{\ParametersChar}}

\global\long\def\ParameterSpace{\Lambda}

\global\long\def\DeltaVar#1{d#1}

\global\long\def\DeltaParameters{\Delta\ParametersChar}

\global\long\def\DerivertiveVar#1{d#1}

\global\long\def\Jacobian{\mathbf{J}}

\global\long\def\PixelScalingVec{\mathbf{a}}

\global\long\def\PixelShiftVec{\mathbf{c}}

\global\long\def\ProjectionOne{p}
 \global\long\def\ProjectionTwo{q}
 \global\long\def\ProjectionThree{r}

\global\long\def\ProjectedPointHx{p}
 \global\long\def\ProjectedPointHy{q}
 \global\long\def\ProjectedPointHz{r}

\global\long\def\RotationLinearChar{\omega}
 \global\long\def\RotationLinearVec{\boldsymbol{\RotationLinearChar}}
 \global\long\def\Expectation{\textrm{E}}

\global\long\def\ErrorSensitivitySet{E}
 \global\long\def\ErrorSensitivitySetFull{\ErrorSensitivitySet_{\mathrm{Full}}}
 \global\long\def\ErrorSensitivitySetRecycle{\ErrorSensitivitySet_{\mathrm{Recycle}}}
 \global\long\def\ErrorSensitivitySetSPAAM{\ErrorSensitivitySet_{\mathrm{SPAAM}}}
 \global\long\def\FocalVector{\mathbf{f}}

\global\long\def\FocalLength{f}

\global\long\def\FocalLengthX{\FocalLength_{x}}

\global\long\def\FocalLengthY{\FocalLength_{y}}

\global\long\def\ImageCenter{p}

\global\long\def\ImageCenterX{\ImageCenter_{x}}

\global\long\def\ImageCenterY{\ImageCenter_{y}}

\global\long\def\IntrinsicFunction{K}

\IEEEdisplaynontitleabstractindextext

%
\IEEEpeerreviewmaketitle

\ifCLASSOPTIONcompsoc
\IEEEraisesectionheading{\section{Introduction}\label{sec:intro}}
\else
\section{Introduction}
\label{sec:intro}
\fi

%
%
%
%

\IEEEPARstart{A}{ugmented} reality (AR) is an interactive, real-time technology, which gives the user the sense that virtual objects exist among real objects, in the physical world.  For example, the user might see a virtual glass sitting next to a physical glass on a tabletop.  A major goal of AR is for the location of the virtual glass to appear as equally real, solid, and believable as the physical one.


In this paper, we refer to this concept as \emph{locational realism}.  We contrast locational realism with the better-known term \emph{photorealism}, which is the traditional computer graphics goal of rendering objects and scenes that are visually indistinguishable from reality.  In AR, the primary goal may not be to render the glass photorealistically, but we are usually interested in the locational realism of the glass---while it may obviously be a cartoon glass, with incorrect illumination and color, we still want its location to be perceived in a manner that is indistinguishable from the location of the physical glass. 


In order for any degree of locational realism to be possible, the AR system must know the 6-degree-of-freedom (6DoF) \emph{pose}---the position ($x$, $y$, $z$) and orientation (\emph{roll}, \emph{pitch}, \emph{yaw})---of the virtual rendering camera within the physical world.  From this information, the system can determine which 2D screen pixels will be required to display a virtual object at a corresponding 3D location (Robinett and Holloway~\cite{robinett1995}).  The more accurately this pose can be known, the greater the locational realism.  


The rendering camera's pose is typically measured using a \emph{tracking system}, which needs to be calibrated in order to report accurate pose estimates. It is possible for the tracking system to directly use a physical video camera within the AR system (Kato and Billinghurst~\cite{kato1999}); otherwise, the tracking system tracks a fiducial that is attached to the AR system.  In this latter case, even though the tracking system needs to report the pose of the AR system's rendering camera, the tracker instead reports the pose of the fiducial.  This leads to the additional requirement that a secondary \emph{calibration} needs to be performed, which yields the transformation between the tracked fiducial and the rendering camera. 


In addition, there are two major ways of displaying AR content.  In video see-though AR (VST AR), the user sees the physical world mediated through a video camera within the AR system.  The system receives a constant stream of image frames from the real world, and combines virtual content to these frames.  VST AR can be used with standard video monitors, handheld devices such as tablets or phones, as well as opaque, VR-style head-worn displays, also referred to as Mixed Reality (MR) displays.  In contrast, optical see-through AR (OST AR) gives the user a view of the physical world directly, while virtual objects are simultaneously imposed into the user's view through optical combiners.  OST AR is almost always accomplished through a transparent head-worn display; although microscopes (Edwards \etal~\cite{edwards2000}) and other optical devices 
are also possible.  While both forms of AR have their respective strengths (and weaknesses), as well as various applications (Billinghurst \etal~\cite{billinghurst2014}), this paper focuses on OST AR. 


Although in VST AR it is possible to use a single camera for both the video stream and the tracking camera (Kato and Billinghurst~\cite{kato1999}), this is never possible in OST AR, because the ``video stream'' comes from the user's eye.  Instead, in OST AR the pose of the head-worn display is tracked, and the AR system needs to know the transformation between the display and the user's eyes.  Therefore, in OST AR a calibration procedure is always necessary. 


This paper surveys and summarizes calibration procedures published until September 2017.  
First, it provides an overview of calibration fundamentals for head-mounted OST AR displays. It then presents an overview of calibration methods, which are categorized according to \emph{manual}, \emph{semi-automatic}, and \emph{automatic} approaches.  Next, it discusses how these calibration methods have been evaluated as well as the metrics used for analysis. Finally, the paper concludes by discussing opportunities for future research.



\section{Fundamentals}
\label{sec:fundamentals}


\global\long\def\CoordinateA{A}

\global\long\def\CoordinateB{B}

\global\long\def\ThreeDPointA{\OperatorA{\ThreeDPoint}{\CoordinateA}}

\global\long\def\ThreeDPointB{\OperatorA{\ThreeDPoint}{\CoordinateB}}

\global\long\def\ScreenPoint{\OperatorA{\PixelPoint}{\screenCoordinate}}

\begin{figure}[t]
\begin{centering}
\includegraphics[width=\columnwidth]{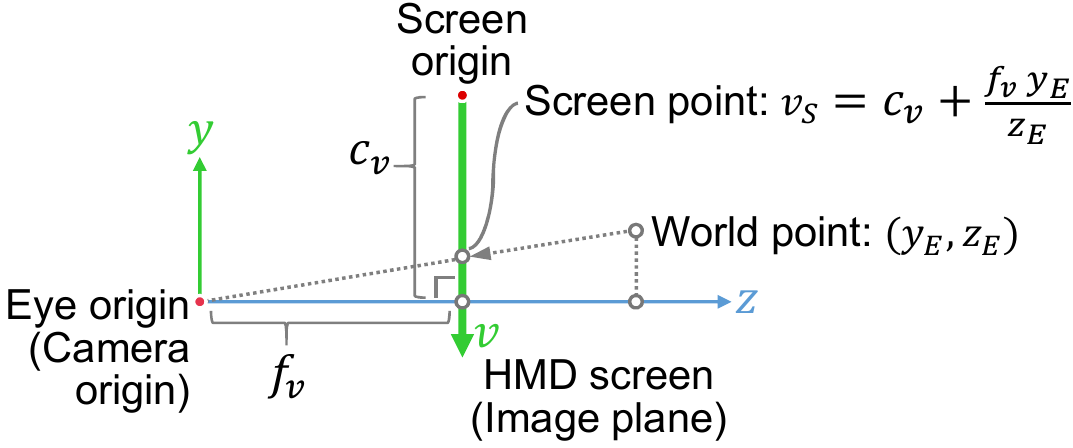} 
\par\end{centering}
\caption{\label{fig:pinhole_camera}The y-z plane of an off-axis pinhole camera model. See also Fig.~\ref{fig:indica}.}
\end{figure}

\subsection{Nomenclature}

Through the paper, we use the following nomenclature. Lower-case letters
denote scalar values, such as a focal length $f_{u}$. Upper-case letters
denote coordinate systems, such as an eye coordinate frame $\eyeCoordinate$.
Lower-case bold letters denote vectors, such as a 3D point in eye
coordinates $\ThreeDpointInE\in\Real^{3}$, or a 2D image point $\PixelPoint\in\Real^{2}$.
Upper-case typewriter letters denote matrices, such as a rotation matrix
$\Rotation\in\Real^{3\times3}$.

We now define a 6DoF transformation from one
coordinate system to another. Given coordinate systems $\CoordinateA$
and $\CoordinateB$, we define the transformation from $\CoordinateA$
to $\CoordinateB$ by $(\RotationAB AB,\TransAB AB)$, where $\RotationAB AB$ is a rotation matrix, and $\TransAB AB$ is a translation vector.  For example,
we can transform $\ThreeDPointA$, a 3D point in $\CoordinateA$,
from $\CoordinateA$ to $\CoordinateB$ by
\begin{equation}
\ThreeDPointB=\RotationAB{\CoordinateA}{\CoordinateB}\ThreeDPointA+\TransAB{\CoordinateA}{\CoordinateB}.
\end{equation}

\subsection{The Off-axis Pinhole Camera Model}

In computer vision, the \emph{intrinsic matrix} $\Intrinsic\in\Real^{3\times3}$ defines the projection transformation from 3D to 2D coordinate spaces. The elements of this matrix describe the properties of the pinhole camera, and its derivation is well described in a plethora of academic texts and research publications \cite{hartley2005multiple,faugeras1993three,hughes2013computer,ma2012invitation,tuceryan2000single,vince2013mathematics}.

Readers desiring to gain a complete and thorough understanding of the physical and mathematical principles behind projection, transformation, or computer graphics in general are encouraged to read the cited publications. Nevertheless, here we provide a brief overview, with the goal of enhancing the reader's understanding of the eye-HMD transformation. 

\begin{figure}[t]
\begin{centering}
\includegraphics[width=1.0\columnwidth]{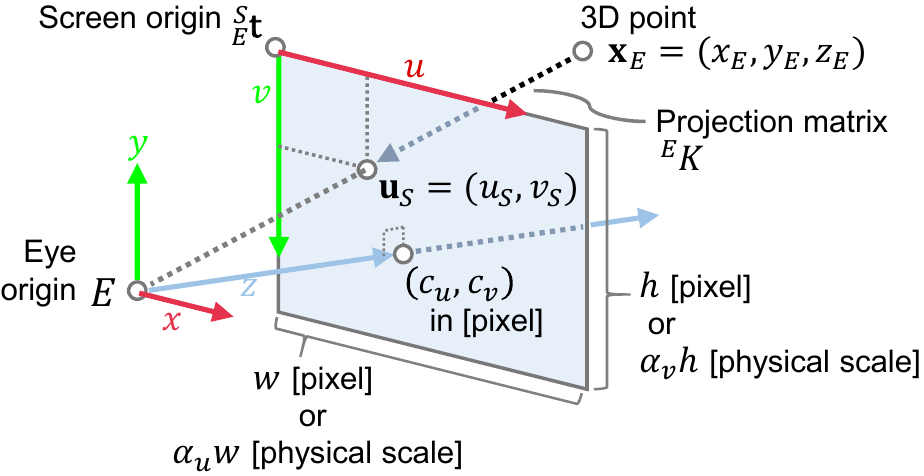} 
\par\end{centering}
\caption{\label{fig:indica}A 3D representation of the image plane, and related intrinsic properties of the pinhole camera model.}
\end{figure}

The eye-HMD system is commonly modeled as an off-axis pinhole camera. 
We define its intrinsic matrix as:
\begin{equation}
\IntrinsicEye=\begin{bmatrix}f_{u} & 0 & c_{u}\\
0 & f_{v} & c_{v}\\
0 & 0 & 1
\end{bmatrix}.\label{eq:ProjectionMatrix1}
\end{equation}
The parameters of $\IntrinsicEye$ are derived directly from the pinhole camera model illustrated in Figures \ref{fig:pinhole_camera} and \ref{fig:indica}. The focal distances $f_{u}$ and $f_{v}$ denote the distances between the imaging plane and the camera center. In the ideal pinhole camera model, the $f_{u}$ and $f_{v}$ components from Equation~(\ref{eq:ProjectionMatrix1}) are identical, meaning the pixels of the image are perfectly square.

For example, given a 3D point in the eye coordinate system $\ThreeDpointInE$, the point is projected to a 2D point $\ScreenPoint$ in the HMD screen space $\screenCoordinate$ by
\begin{eqnarray}
\ScreenPoint = \IntrinsicEye\ThreeDpointInE.
\end{eqnarray}

In practice, however, we first obtain $\ThreeDpointInE$ as the 3D point $\ThreeDpointInW$, in the HMD coordinate system\footnote{%
The HMD coordinate system is typically defined by an inside-out looking camera or an outside-in looking tracking system that determines the pose of a fiducual}. 
Therefore, we first transform $\ThreeDpointInW$ into $\ThreeDpointInE$ by
\begin{equation}
\ThreeDpointInE=\RotationFromWorldToEye\ThreeDpointInW+\TransFromWorldToEye,
\end{equation}
where the rotation matrix $\RotationFromWorldToEye\in\Real^{3\times3}$, and the translation vector $\TransFromWorldToEye\in\Real^{3}$, represent a transformation from the HMD coordinate system $\worldCoordinate$, which is attached to the HMD, to the user's eye's coordinate system $\eyeCoordinate$.

\begin{figure*}[t]
\includegraphics{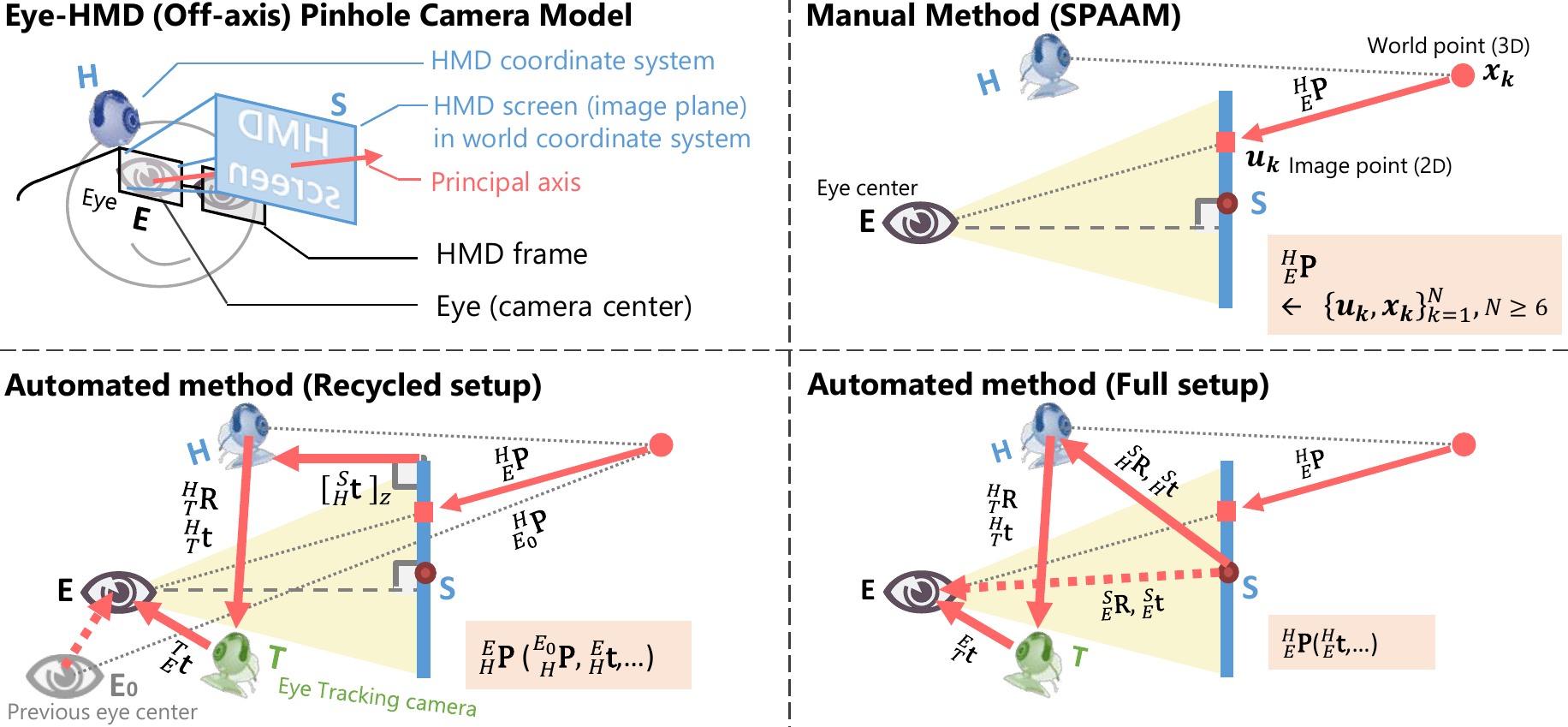}\caption{Illustration of pinhole camera projection. (top left) coordinate systems of an eye-HMD system. (top right) Manual method. (bottom left) Automated method with the recycled setup. (bottom right) Automated method with the full setup.}
\label{fig:projection} 
\end{figure*}

By integrating this transformation into the camera model $\IntrinsicEye$, we obtain a $3\times4$ projection matrix $\ProjectionWorldToEye$, from the display (HMD) coordinates to the user's eye's coordinates: 
\begin{equation}
\ProjectionWorldToEye=\IntrinsicEye\left[\begin{array}{cc}
\RotationFromWorldToEye & \TransFromWorldToEye\end{array}\right]\in\Real^{3\times4}\label{eq:Camera1}.
\end{equation}
Figure~\ref{fig:projection}, top left, is another illustration of these coordinate systems. 

Therefore, all calibration methods must be able to produce $\ProjectionWorldToEye$, either by solving for all of the matrix components at once, or by systematically determining the parameters in Equation~(\ref{eq:Camera1}).

Generally, when solving for all of the components of $\ProjectionWorldToEye$ at once, the most common approach is the direct linear transformation (DLT)~\cite{abdel1971direct,sutherland1974dlt,hartley2003multiple}. This method estimates $\ProjectionWorldToEye$ by solving a linear equation, which is constructed from a minimum of 6 3D-2D correspondences.  Given the linear solution as an initial estimate, a non-linear optimization method, such as Levenberg-Marquardt \cite{more1978levenberg,hartley2003multiple}, can then be applied. 


\subsection{Modeling the Intrinsic Matrix of OST HMDs}

While most rendering engines used for computer graphics presume $\ProjectionWorldToEye$ as an ideal pinhole
camera, in physical implementations, this model may be unequal as
a result of distortion, imperfections on the imaging plane, non-uniform
image scale, etc.. In this case an alternative model using a single
focal length value and the image aspect ratio may be more appropriate~\cite{forsythbook}.

The ``principal axis'', in this alternative, lies perpendicular to the imaging plane and
extends to the aperture. The intersection of the principal axis and
the imaging plane occurs at the ``principal point''. Ideally, the
principal point would occur at the origin of the image coordinate
system. However, when this is not the case, the parameters $c_{u}$
and $c_{v}$, illustrated in Figure~\ref{fig:indica}, represent the offset from the origin. $\tau$ represents
a skew factor when the axes of the image plane are not orthogonal,
which would produce an image plane resembling a parallelogram instead
of a rectangle or square.

When the camera is located at, and is orthogonal to, the origin of
the 3D coordinate space, then the transformation of objects into the
camera frame of reference is implicit. However, should the camera
move to another viewing location in the world, as is often the case,
then an extrinsic transformation is required to transform the coordinates
of the objects in the world into the camera frame. This transform
is the $(\RotationFromWorldToEye,\TransFromWorldToEye)$ component
of Equation~(\ref{eq:Camera1}), referred to as the extrinsic component.


The $\RotationFromWorldToEye$ describes the rotation of the camera
with respect to the world coordinate axes and the $\TransFromWorldToEye$
denotes the translational offset from the origin along the X, Y, Z
cardinal directions. This transformation, with respect to OST HMD
calibration, represents the transformation of the tracked coordinate
frame of the HMD relative to the eye.
Unfortunately, the location of the user's optical center, or
alternatively the nodal point, is not easily determined at run-time. Nonetheless, given the extrinsic and intrinsic parameters, calculation
of the 12 values in the final camera projection matrix $\ProjectionWorldToEye$
in Equation~(\ref{eq:Camera1}) is achieved through simple matrix multiplication.

In an OST HMD system, we can further break down the intrinsic matrix
by using the position of the virtual screen of an OST HMD with respect
to the eye.

Given an eye tracker $\cameraCoordinate$ attached on an HMD with
a pose $(\RotationFromWorldToCamera,\TransFromWorldToCamera)$ from
the HMD to the tracking camera, we can get the position of the eye
with respect to the screen as $\TransFromScreenToEye$ if we also
know the pose of the screen $(\RotationFromScreenToWorld,\TransFromScreenToWorld)$
(Figure~\ref{fig:indica} and Figure~\ref{fig:projection} bottom right).

Assuming that we know that position as the translation
vector $\TransFromScreenToEye=[\screenEyeCenterX,\screenEyeCenterY,\screenEyeCenterZ]^{\Transpose}$
then the intrinsic matrix in Equation~(\ref{eq:ProjectionMatrix1})
can be defined as the following \cite{itoh2014interaction,itoh2014performance}
(Figure~\ref{fig:indica} and Figure~\ref{fig:projection} bottom right):
\begin{align}
\IntrinsicEye & =\left[\begin{array}{ccc}
\pixelScaling_{u} &  & 0\\
 & \pixelScaling_{v} & 0\\
 &  & 1
\end{array}\right]\left[\begin{array}{ccc}
\screenEyeCenterZ &  & -\screenEyeCenterX\\
 & \screenEyeCenterZ & -\screenEyeCenterY\\
 &  & 1
\end{array}\right],\label{eq:FullSetup}
\end{align}
where $\screenCoordinate$ is the virtual screen coordinate system
and $\PixelScalingVec=[\pixelScaling_{u},\pixelScaling_{v}]^{\Transpose}$
is the scaling factor that converts 3D points on the screen to pixel
points.

We call this formulation the \textit{full setup}, since it constructs the intrinsic matrix from explicit display parameters).
Note that the formulation in \cite{itoh2014interaction,itoh2014performance} defines $\TransFromScreenToEye$
at the center of the screen, whereas our formulation defines it at the screen origin,
which makes the formulation simpler.

Note that $\TransFromScreenToEye$ is dependent
on the current position of the user's eye with respect to the display,
thus the intrinsic matrix varies when the display is repositioned
on one's head.

If we know an old intrinsic matrix $\IntrinsicSPAAM$ based on an old eye position  
$\TransFromScreenToEyeOld=[\screenEyeCenterX_0,\screenEyeCenterY_0,\screenEyeCenterZ_0]^{\Transpose}$, 
\begin{align}
\IntrinsicSPAAM & =\left[\begin{array}{ccc}
\pixelScaling_{u} &  & 0\\
 & \pixelScaling_{v} & 0\\
 &  & 1
\end{array}\right]\left[\begin{array}{ccc}
\screenEyeCenterZ_0 &  & -\screenEyeCenterX_0\\
 & \screenEyeCenterZ_0 & -\screenEyeCenterY_0\\
 &  & 1
\end{array}\right],\label{eq:FullSetupOld}
\end{align}
we can \textit{update} it to the new intrinsic matrix $\IntrinsicEye$ as follows (Figure~\ref{fig:projection} bottom left):
\begin{align}
\IntrinsicEye & =\IntrinsicSPAAM\left[\begin{array}{ccc}
1+\TransFromEyeToSPAAMsZ/\screenEyeCenterZ_0 &  & -\TransFromEyeToSPAAMsX/\screenEyeCenterZ_0\\
 & 1+\TransFromEyeToSPAAMsZ/\screenEyeCenterZ_0 & -\TransFromEyeToSPAAMsY/\screenEyeCenterZ_0\\
 &  & 1
\end{array}\right],\label{eq:RecycleSetup}
\end{align}
where $[\TransFromEyeToSPAAMsX,\TransFromEyeToSPAAMsY,\TransFromEyeToSPAAMsZ]^{\Transpose}=\TransFromScreenToEye-\TransFromScreenToEyeOld$
is a translation from the old eye position $\TransFromScreenToEyeOld$ to the new eye position $\TransFromScreenToEye$, 
in other words $\TransFromSPAAMToEye$ by our notation convention. We call this as the \textit{recycled setup} compared to the full setup.

Equation \ref{eq:FullSetup} (full setup) does not rely on knowledge
about a previous eye position $\TransFromWorldToSPAAMs$. Instead,
it requires the{} virtual screen pose $(\RotationFromWorldToScreen,\TransFromWorldToScreen)$
and the scaling vector $\PixelScalingVec$ $[\mathrm{pixel}/\mathrm{meter}]$.
On the other hand, Eq. \ref{eq:RecycleSetup} (recycled setup) does
not rely on these parameters, except for $\left[\TransFromWorldToScreen\right]_{z}$,
because it reuses the old intrinsic matrix $\IntrinsicSPAAM$.



\subsection{Estimating Projection}

Unfortunately, it is rarely, if ever, possible to explicitly possess the exact intrinsic and extrinsic parameters for a specific HMD and user configuration at run time. Therefore, OST HMD calibration procedures often utilize manual user interaction techniques in order to produce an approximation, or estimate, of the final projection matrix parameters. Initial manual calibration modalities, for example, adapt existing computer vision camera calibration mechanisms, which utilize pixel to world correspondences for determining the viewing parameters. These adapted techniques do not obtain all correspondences at once, as would be possible in an image captured from a camera, but instead reduce the strategy to simple bore-sighting through which each separate correspondence is recorded in sequence~\cite{caudell1992augmented, klinker1999augmented}. The correspondence data obtained from this process includes both the 2D pixel location of the on--screen reticle and the 3D location of the physical alignment point. Values from multiple alignments can then be combined into a system of linear equations describing the projection of the 3D point into the 2D space and solved using standard methods 
in the context of DLT discussed above. The solution to this linear equation system is the complete set of parameters describing the projection matrix, or virtual camera, from Equation~(\ref{eq:Camera1}).

The bore--sighting schema though, forces a number of requirements, including placement of the HMD such that the user's view is perpendicular to the display screen and that the user is able to reliably align the on-screen indicator with a high level of precision. In order to satisfy these conditions, the user's head must be rigidly secured, preventing movements which may shift the display screen or disrupt the alignment process. Inhibition of user movement makes this methodology not only uncomfortable and tedious, but also impractical for use outside of a laboratory setting. Successive iterations and adaptations have fortunately enabled a relaxation of the fixation constraint by affording a compromise with other requirements as well.

Within the next sections we will discuss the evolution of approaches targeted at estimating these intrinsic and extrinsic parameters, as well as methods which propose calibration models which diverge from the pinhole camera model.

\section{Manual Calibration Methods\label{sec:manual_calibration}}

\begin{figure*}[t!]
    \center
    \includegraphics[width=0.9\textwidth]{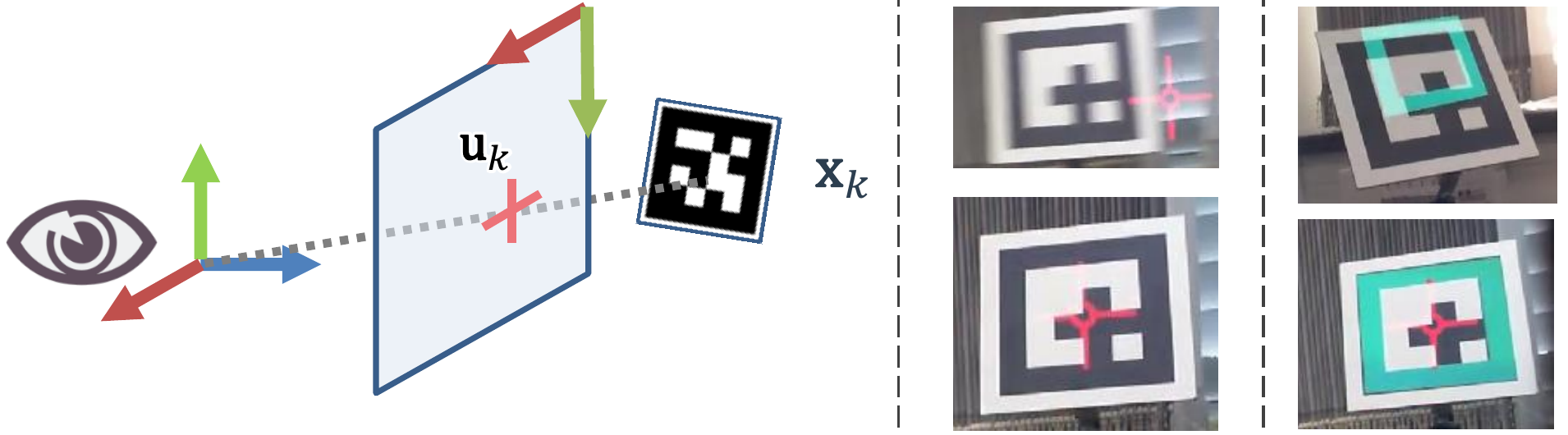}
    \caption{Data collection in SPAAM. Left: A single 2D point $u_k$ is manually aligned with a 3D point $x_k$. Middle: Ego-centric view through an OST HMD aligning a virtual 2D cross hair with a 3D tracked marker. Right: Green virtual square overlaid on the physical marker before and after the calibration.}
    \label{fig:spaam}
\end{figure*}

\begin{figure*}
\includegraphics[width=0.98\textwidth]{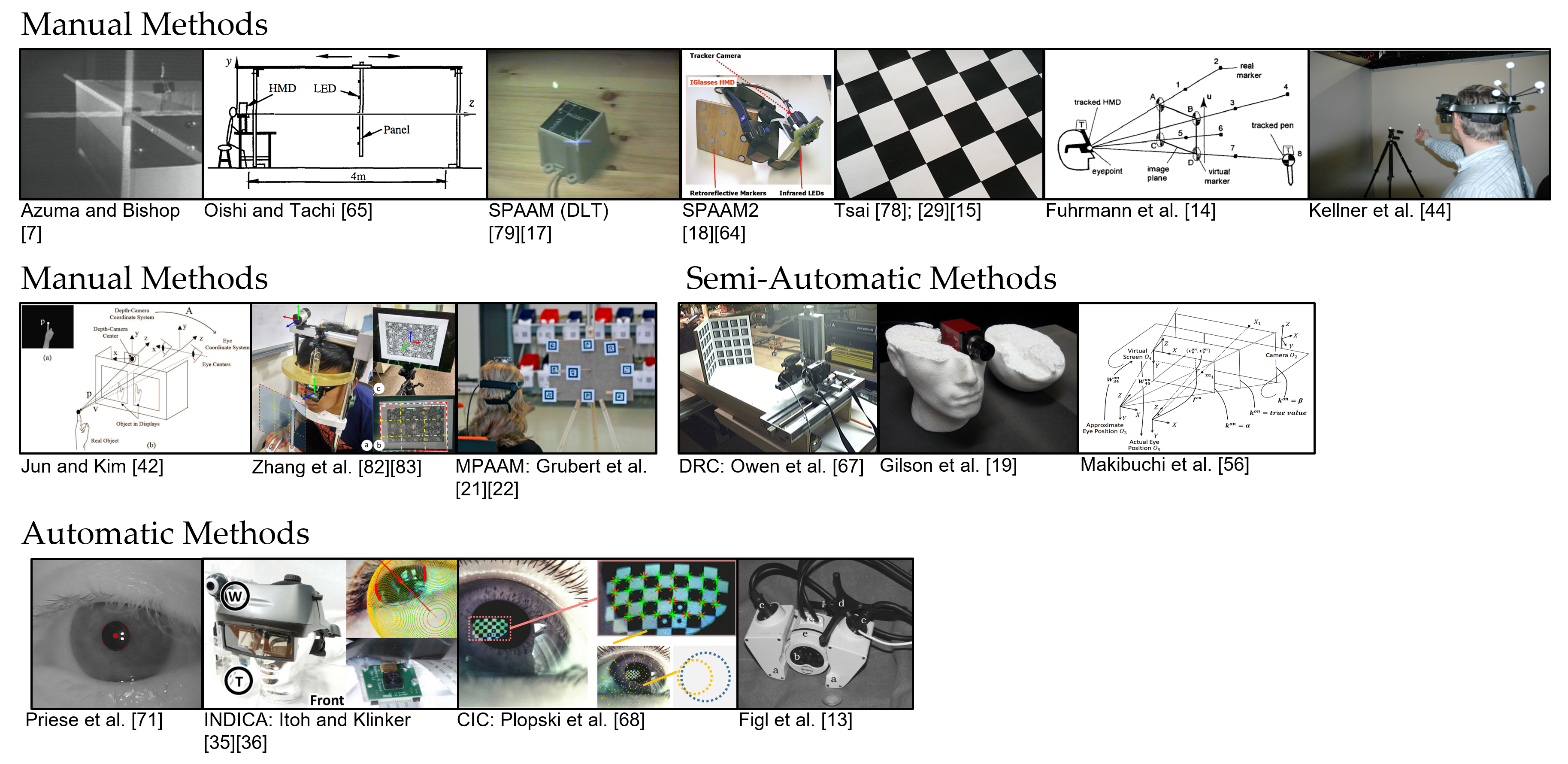}\caption{Thumbnail images from the rows of  Table~\ref{tab:calib}.}
\label{fig:thumbs}
\end{figure*}

This section summarizes the methods where the calibration requires the human operator to perform manual tasks.  The top portion of Table~\ref{tab:calib} summarizes these methods, and for each method Figure~\ref{fig:thumbs} gives a key thumbnail image. 





Calibration through parameter estimation has yielded a number of procedures that rely upon user interaction to collect 3D-2D correspondences by manually aligning a world reference point to 2D points displayed on the screen of an OST HMD.  Azuma and Bishop~\cite{azuma1994improving} propose estimating the extrinsic virtual camera parameters by manually aligning  virtual squares and a cross-hair with a wooden box. For estimating the field-of-view of the virtual camera the user must  simultaneously align two virtual lines with the box edges.

Janin \etal \cite{janin1993calibration} proposed two methods to estimate parameters associated with the external sensor, a virtual screen and user-specific parameters (eye location): direct measure of the parameters and optimization-based. As the authors note, it is difficult to accurately measure relevant parameters, and, hence recommend parameter estimation through optimization. Similar, to Azuma and Bishop~\cite{azuma1994improving} they propose to align a virtual cursor with a physical registration device with known geometry. The authors note, that the joint estimation of the 17 parameters of their model is susceptible to noise but do not quantify their calibration results.

Oishi et al.~\cite{oishi1996methods} use an elaborate "shooting gallery" calibration setup, which presents LEDs fixed on a large plate 0.5 to 4 meters from the user, and a predetermined virtual projection model, which matches the physical calibration environment and the physical HMD. The head of the user has to be fixed during the procedure in the world coordinate origin, and virtual points have to be aligned with physical targets manually. The calibration process has the following steps:
First, the HMD is positioned at the world coordinate origin. Then a physical calibration pattern has to be matched with a virtual target indicator (using a control joystick, 13 times per eye). If the recorded matches are below a predetermined threshold the system is calibrated. If the mismatch is too large, further correspondences are collected and the projection model is updated.

Tuceryan and Navab~\cite{tuceryan2002single} introduced SPAAM (Single Point Active Alignment Method), as an improvement to the manual alignment data collection scheme (Figure \ref{fig:spaam}). They propose collecting individual 2D-3D point correspondences, one at a time, and then solving for all projection parameters simultaneously. To this end, the user is asked to align a single 2D symbol (e.g., cross or circle) with a 3D object. Both the HMD and the 3D object are spatially tracked. After collecting sufficient correspondences (at least 6), the correspondences are used to create and solve a system of linear equations according to the DLT method introduced in section \ref{sec:fundamentals} \cite{abdel1971direct, sutherland1974dlt, hartley2003multiple}.
The biggest advantage of SPAAM is its weak requirements of hardware: needing only a tracking system for calibration to be done by a user ad-hoc. Thus SPAAM can easily be integrated into most OST HMD applications.

Unfortunately, the manual procedure inevitably induces human related errors during the data collection, due to imperfect alignments from user posture~\cite{axholt2008} and input actions \cite{maier2011empiric}. Furthermore, these errors may be sufficiently high, for users not familiar with SPAAM, to render the calibration a failure, requiring the need to repeat the procedure multiple times. 
Despite the potential drawbacks, SPAAM has proved to be a popular and influential calibration method, onto which a number of improvements to the original approach have been proposed.

Instead of performing a completely new calibration every time a user puts on an HMD, Genc et al.~\cite{genc2002practical} proposed Two-Stage SPAAM (SPAAM2), which reuses existing calibrations.  Their process works as follows: Initially, all 11 parameters (extrinsics + intrinsics) are estimated. If the user removes the HMD and then later puts it back on, only a subset of those parameters are re-estimated. The intrinsics of the virtual camera are assumed to not change over time, only the position of the virtual camera center (i.e., the position of the eye's nodal point relative to the display screen). Therefore, linear scale and shift parameters are estimated, which correct for a potential image shift and scale change due to the new projection center. 

Using their updated model, the user only needs to collect two point correspondences. However, the justification of SPAAM2 is that the 3D shift of the virtual camera center can be modeled by a linear scale and shift transformation on the image plane. 
Their assumption is rather "redefining" the scale and position of the display's image plane under the assumption that the orientation of the plane stays the same.  We elaborate the theory behind this in Section \ref{sec:auto}.



When using a vision-based inside-out camera system for 3D tracking, 
Genc et al.~\cite{genc2001optical} also propose to avoid computing the pose between the external camera system and 3D object directly.  Instead, and under the assumption that the camera and HMD are rigidly attached to each other, they present a formulation that uses the projection matrix of the inside-out tracking camera to estimate the projection matrix of the virtual camera. Unfortunately, they do not present results that are significantly better than the base algorithm (SPAAM + explicit pose computation). 

 Fuhrmann et al.~\cite{fuhrmann1999fast} propose to determine the parameters of the virtual viewing frustum by collecting 8 2D-3D point correspondences per eye that define the viewing frustum corner points. They propose to further reduce the number of needed point correspondences to two per eye for adopting the projection for individual users. This can be achieved under the assumption that the projection of the 3D point intersects the virtual image plane at a known distance. Now, only the eye position has to be determined and the user only needs to provide point samples for two opposite corners of the display. For distortion correction, the authors propose to fallback on a camera-based detection of a distorted line pattern (or alternatively let the user specify many points of intersecting lines).
 
 Another data collection scheme was proposed by Kellner et al.~\cite{kellner2012geometric} in 2012. They propose to "aim" at a distant 3D target with another handheld target, resulting in 2D point-to-3D line correspondences.  Subsequently, they first determine the display rotation and translation, and then the focal length and principal point. While the method results in a shorter acquisition time compared to SPAAM, it also results in larger calibration errors.
 
 Instead of aiming with the head, several approaches proposed to move a handheld target instead.  O'Loughlin and Sandor~\cite{o2013user} proposed to use a handheld marker for alignment, and Moser et al.~\cite{moser2016evaluation} investigated the contextual impact of user-centric tracking markers, finding that simple stylus alignment is preferable to finger tracking for 3D point input.  The latter approach is also employed in the calibration of the Microsoft HoloLens HMD\footnote{Please note, that the Microsoft HoloLens does not offer a complete user-based OST calibration procedure, but solely determines the interpupilary distance - see https://developer.microsoft.com/en-us/windows/mixed-reality/calibration - last accessed September 2nd, 2017.}.

In contrast to SPAAM, Grubert et al.~\cite{grubert2010comparative, grubert2008untersuchungen}, propose to collect multiple 2D-3D point correspondences simultaneously in a Multiple Point Active Alignment scheme. Here, the user aligns a grid of 9 3D points aligned on calibration board placed at a distance of about 150 cm. While the calibration procedure significantly speeds up the data collection phase, it also results in larger calibration errors. 

The manual calibration techniques for Optical See-Through calibration has also seen application in other HMD domains types, particularly to head-mounted projective displays (HMPDs). Hua et al. developed a HMPD and accompanying calibration procedures \cite{hua2002calibration,gao2003easy}. Their calibration relies on Tsai's calibration technique \cite{tsai1987versatile}. Hence, for data collection, they show a printed grid 14x13, on which the user has to align a virtual cross. This procedure is repeated at least two times. The biggest disadvantage of using the Tsai method is that, in practice, a large number of point correspondences have to be collected and that for at least two grid poses. For example, in the work of Hua et al. the authors conducted the calibration with 10 different grid poses resulting in 14x13x10 = 1820 correspondences, which needed to be aligned \cite{hua2002calibration}. The authors argue, that the data collection could be automated by placing a camera in the exit pupil \cite{gao2003easy}. However, this could lead to additional errors as the camera position during calibration is not identical with the eye position during use. SPAAM was also used for other custom HMD designs, such as \cite{zhang20153d}, in which a natural feature tracking target was used for collecting 27 point correspondences.

In 2016, Jun and Kim \cite{jun2016calibration} proposed a calibration method for stereo OST-HMDs equipped with a depth camera. They presented a simplified HMD-eye model assuming collimated displays with no focal length and perceptual pinhole centers for the eyes (i.e. the perceptual eye projection). Their model solves for the extrinsic parameters of the depth camera, the interpupillary distance of the user and the position of the users' eyes. The authors claim that a full calibration can be achieved with 10 point correspondences (collected by pointing with a finger on a 2D circle). After initial full calibration, only the user parameters (interpupillary distance, eye position) are estimated in subsequent simplified calibrations.

In 2017, Zhang \etal \cite{zhang2017ride, zhang2017accurate} proposed a dynamic SPAAM method that considers eye orientation to optimize the projection model. Their method, RIDE (region-induced data enhancement), splits the user's FoV into 3-by-3 segments and update the main projection matrix to adapt the shift of the eye center (the nodal point) due to eye orientation.

While manual calibration methods can achieve accurate results, the burden on users in terms of time and workload can be substantial. Empirically, we found that many users would calibrate an HMD (at most) once per work session or when the display is used for the first time. The calibration process is of an open-loop nature, requiring substantial effort from the users to successfully complete the calibration task. Hence, the need for automated, closed-loop methods arises, which will be discussed in the following section.


\begin{table*}[!th]
\caption{Overview of calibration methods. \textbf{MC}: minimum number of 2D\textendash 3D correspondences. \textbf{Parameters}: estimated parameters. \textbf{Alignment Mode}: \textbf{FIX}: fixed head or camera-rig, \textbf{H}: through head-movement, \textbf{F}: through finger or hand movement. \textbf{Data Collection}: \textbf{i}: individually (1 correspondence at a time), \textbf{m}: multiple correspondence at once. 
Figure \ref{fig:thumbs} shows representative thumbnail images from each method. 
\vspace{-3.5mm}
}
\label{tab:calib}
\begin{center}
\begin{tabular}{c>{\centering}p{0.17\textwidth}>{\centering}p{0.11\paperwidth}>{\centering}p{0.2\paperwidth}c>{\centering}p{0.13\paperwidth}}
\toprule
 & \textbf{Method} & \textbf{MC} & \textbf{Parameters} & \textbf{Alignment Mode} & \textbf{Data Collection}\tabularnewline
\toprule
\multirow{8}{*}{\begin{turn}{90}
\textbf{Manual Methods}\hspace{1cm}
\end{turn}} & Azuma and Bishop \cite{azuma1994improving}  & 8 points + 4 lines  & eye location, FoV  & H  & points i, lines m (2) \tabularnewline
\cmidrule(l){2-6}
 & Oishi and Tachi \cite{oishi1996methods}  & 13 per eye  & eye location  & FIX  & i\tabularnewline
\cmidrule(l){2-6}
 & SPAAM (DLT): \\Tuceryan \etal\cite{tuceryan2002single} \\Genc \etal\cite{genc2001optical}  & min 6  & projection matrix (full)  & H (F in \cite{o2013user,moser2016evaluation})  & i \tabularnewline
\cmidrule(l){2-6} 
 & SPAAM2 / EasySPAAM: \\Genc \etal\cite{genc2002practical} \\Navab \etal\cite{navab2004line}  & 2  & scale, shift  & H  & i \tabularnewline
\cmidrule(l){2-6} 
 & Tsai \cite{tsai1987versatile}: \\Hua \etal\cite{hua2002calibration} \\Gao \etal\cite{gao2003easy}  & 14$\times$13 $\times$ (2..10) = 364\textasciitilde{}1820 \cite{hua2002calibration} & all intrinsics + extrinsics  & FIX \cite{gao2003easy} or F \cite{hua2002calibration}  & m \cite{gao2003easy} or i \cite{hua2002calibration}\tabularnewline
\cmidrule(l){2-6} 
 & Fuhrmann \etal\cite{fuhrmann1999fast}  & 8 (full) 2 (update)  & all intrinsics + extrinsics (full), \\eye position (update)  & F  & i \tabularnewline
\cmidrule(l){2-6} 
 & Kellner \etal\cite{kellner2012geometric}  & 5  & all intrinsics + extrinsics  & H+F  & i \tabularnewline
\cmidrule(l){2-6} 
 & Jun and Kim \cite{jun2016calibration}  & 3 (full), 2 (update)  (10 and 5 recommended) & offline only: extrinsic orientation of depth camera, interpupillary distance, eye positions & F  & i \tabularnewline
 \cmidrule(l){2-6} 
 & Zhang \etal \cite{zhang2017ride, zhang2017accurate}  & 9 regions x 6 samples = 54 (3x3 grid with standard SPAAM) & multiple projection matrices for individual regions (full), single projection matrix for update & H  & i \tabularnewline
 \cmidrule(l){2-6} 
 & MPAAM: \\Grubert \etal\cite{grubert2010comparative}, \cite{grubert2008untersuchungen}  & 6  & projection matrix (full)  & H  & m \tabularnewline
\midrule
\multirow{2}{*}{\begin{turn}{90}
\begin{tabular}{c}
\textbf{Semi-}\tabularnewline
\textbf{automatic}\tabularnewline
\end{tabular}
\end{turn}} & DRC: \\Owen \etal\cite{owen2004display}  & offline: 20 \\online: 1  & offline: all intrinsics + extrinsics + radial distortion + spherical aberration. online: eye location, focal length  & offline: FIX, online: H  & offline: m, online: i \tabularnewline
\cmidrule(l){2-6} 
 & Gilson \etal\cite{gilson2008spatial}  & 30  & all intrinsics + extrinsic  & FIX  & m \tabularnewline
\cmidrule(l){2-6} 
 & Makibuchi \etal\cite{makibuchi2013vision}  & offline: 4 \\online: 4  & offline: virtual screen pose and approximate eye location. \\online: current eye location  & offline: FIX, online: H & m \tabularnewline
\midrule
\multirow{4}{*}{\begin{turn}{90}
\textbf{Automatic}\hspace{1.0cm}
\end{turn}} & Priese \etal\cite{priese2007automatische} & not applicable (NA)  & eye location  & NA  & NA \tabularnewline
\cmidrule(l){2-6} 
 & INDICA: \\Itoh and Klinker \\\cite{itoh2014interaction}, \cite{itoh2014performance}  & offline: 6 \\online: NA  & offline: projection matrix (full) or virtual screen pose. \\online: eye location  & offline: FIX or H  & offline: i or m, \\online: NA \tabularnewline
\cmidrule(l){2-6} 
 & CIC: \\Plopski \etal\cite{plopski2015corneal}  & online: \\2 $\times$ 3 frames  & offline: display screen pose and eyeball parameters. \\online: eye location  & FIX  & m \tabularnewline
\cmidrule(l){2-6} 
 & Figl \etal\cite{figl2005fully}  & unknown  & offline only: eye location, \\focal length  & FIX  & m \tabularnewline
\bottomrule
\end{tabular}
\end{center}
\end{table*}

\section{Automating Calibration} \label{sec:auto}


Unlike those manual calibrations reviewed in Section~\ref{sec:manual_calibration}, some works propose (semi-)automatic calibration methods. A common idea behind these automated methods is to formulate an OST HMD system as the combination of a display model and an eye model. Indeed, a projection matrix from SPAAM implicitly models the system as a planar display screen floating in midair, with the eye center position relative to the screen.  This leads to an off-axis pinhole camera model as discussed in Section 2. Given an OST HMD model, these automated methods measure the parameters on-line, and/or estimate them prior to the actual calibration. Overall, these methods simplify the calibration procedure.

\subsection{Semi-Automatic Calibration  Methods}

This section introduces methods, which by estimating a subset of parameters in separate calibration stages, attempt to minimize the number of point correspondences that need to be manually collected by the human operator.  The middle portion of Table~\ref{tab:calib} summarizes these methods, and for each method Figure~\ref{fig:thumbs} gives a key thumbnail image. 

In early works, Genc, Tuceryan, and Navab~\cite{genc2002practical} and Navab et al.~\cite{navab2004line} developed the Easy SPAAM method, which updates an old projection matrix from a previous SPAAM calibration using a simple manual adjustment. This method assumes that the matrix change can be modeled with a 2D warping of the screen image, including scaling.  Therefore, fewer parameters are needed, and users need to collect only two or more 2D-3D correspondences.

After Easy SPAAM, Owen et al. \cite{owen2004display} propose Display-Relative Calibration (DRC). Their work is one of the first attempts to explicitly split an OST HMD system into a display model and an eye model. In DRC, the authors proposed a two-step calibration process.  They first create an off-line calibration for the display model using a mechanical jig, and then propose 5 different options for the on-line estimation of the eye model.  The options involve varying degrees of simplifying assumptions, ranging from not performing any on-line calibration, to performing a Easy SPAAM-like simple warping, and finally to a full 6 DoF eye pose estimation.

Similar to Owen et al.\cite{owen2004display}, Gilson et al. \cite{gilson2008spatial} propose replacing the user's eye with a camera and exploit established camera calibration techniques for determining the virtual camera parameters.  They differ from Owen's work in that they do not need measurements conducted by a human operator, instead they take images directly through the HMD optics. They also found that no user adaption was needed for their calibration techniques.
 
In 2013, Makibuchi et al. \cite{makibuchi2013vision} proposed a vision-based robust calibration (ViRC) method.  It first uses a view-point camera for off-line parameter estimation.  Then, a camera attached to the HMD tracks a fiducial marker as the user aligns the marker with a crosshair on the screen.  Using the correspondences, the perspective-n-point algorithm (PnP) optimizes both offline and on-line parameters at the same time.  They find that, compared to the direct linear transform (DLT) method, the ViRC method requires fewer user input trials, and achieves up to 83\% more accurate reprojection errors.

The methods proposed so far, allow human operators to lower the number of point correspondences required for a successful calibration. However, this partially comes at the cost of separate and elaborate calibration phases, which can require additional hardware such as cameras or a calibration rig \cite{owen2004display,gilson2008spatial,makibuchi2013vision}.


\subsection{Automatic Calibration Methods}


Finally, this section covers methods which attempt to completely free the human operator from having to manually perform any calibration procedures.  The bottom portion of Table~\ref{tab:calib} summarizes these methods, and for each method Figure~\ref{fig:thumbs} gives a key thumbnail image. 


Luo et al.~\cite{luo2005registration} developed an on-axis camera for eyeglass-like OST HMDs, which in theory eliminates the need for manual calibration.  However, because of the optical design's small size, the camera must be placed 20 mm behind the user's eye location, which can lead to registration errors at close distances.

In 2007, Priese et al.~\cite{priese2007automatische}, after an initial full calibration, proposed estimating the eye location using eye tracking.  However, they tested their approach only using static images of an eye and did not verify the system with actual users.

Figl et al.~\cite{figl2005fully} presented a method for the determination of focal lengths and eye location for a binocular medical HMD (Varioscope M5), using a fully automated setup, including stepping motors for changing the distance of a calibration pattern.  However, after this initial camera-based calibration, they do not consider the calibration of the user's actual eye position. 

In 2014, Itoh and Klinker~\cite{itoh2014interaction} proposed the INteraction-free DIsplay CAlibration (INDICA) method, which utilizes an eye-tracker installed on an OST HMD.  Their method measures the eye center on-line and automatically generates a projection matrix. They use the same pinhole camera model as SPAAM.  The display parameters are decomposed from a projection matrix, which is obtained from a SPAAM calibration performed once off-line beforehand (Itoh and Klinker~\cite{itoh2014performance}). Their follow-up work evaluated INDICA with display parameters calibrated off-line via a camera, which means the method operates totally without the need of additional user  input~\cite{itoh2014interaction}. 

In Section~\ref{sec:manual_calibration}, we mentioned that the assumption of SPAAM2 (Genc \etal\cite{genc2002practical}) leads to a different interpretation. Based on this assumption, we get
\begin{align}
\ProjectionWorldToEye^{\prime} & =\underbrace{\left[\begin{array}{ccc}
\pixelScaling_{x}^{\prime} &  & \CameraCenter_{x}^{\prime}\\
 & \pixelScaling_{y}^{\prime} & \CameraCenter_{y}^{\prime}\\
 &  & 1
\end{array}\right]}_{\IntrinsicEye^{\prime}}\ProjectionWorldToEye\\
 & =(\IntrinsicEye^{\prime}\IntrinsicEye)\left[\begin{array}{cc}
\RotationFromWorldToEye & \TransFromWorldToEye\end{array}\right],
\end{align}
 where $\IntrinsicEye^{\prime}$ denotes the scale and shift parameters. This means that SPAAM2 redefines the screen parameter matrix as $\IntrinsicEye^{\prime}\IntrinsicEye$.
Since the screen parameters should stay the same, this interpretation
is incorrect. An implicit assumption of SPAAM2 is that only the
eye center position changes, which actually leads to Eq.~\ref{eq:RecycleSetup} instead.
And, the three parameters $\TransFromSPAAMToEye$ could be estimated
via two 2D-3D data correspondences.


\ParLabel{Eye Models}
Plopski et al.~\cite{plopski2015corneal} propose another automated method: Corneal-Imaging Calibration (CIC). Unlike INDICA, which uses an iris-based method for eye-tracking, CIC estimates the eye position by utilizing a reflection of an image on the cornea of a user's eye---an effect known as the corneal reflection. In CIC, a fiducial pattern is displayed on an HMD screen, and an eye camera captures its corneal reflection. CIC then computes the rays that are reflected on the eye cornea and pass through corresponding display pixels. Given the 3D pose of the display in the HMD coordinate system, the diameter of the cornea sphere under the dual circle eye model, and a minimum of two rays, the method computes the position of the corneal sphere of the eyeball. Then, given three corneal sphere positions while the eyeball is rotating, CIC estimates the 3D center of the eyeball.  This eye position estimation, based on the reflected features and a simplified model of the eye’s structure, yields more accurate 3D localization estimates than direct iris detection.

However, the 3D eye model that both INDICA and CIC use can be improved.  The model assumes that the eyeball can be schematically modeled as two intersecting 3D spheres, where the first sphere models the spherical part of the eyeball that consists of the sclera, and the second sphere models the cornea curvature.  Under this model, the optical center of the eye camera is assumed to be located at the center of the sclera (eyeball) sphere.  However, the nodal point of the eye---the point where light rays entering through the pupil will intersect---would be a more appropriate location for the optical eye center (c.f. Jones \etal\cite{jones2016schematic}).


\ParLabel{Display Models}\label{sec:display_model}
Most of the methods we mentioned so far treat the image screen of an OST HMD as a planar panel. However, this model ignores the fact that the combining optics could distort the incoming light rays before they reach the eye, in a manner similar to corrective glasses.  This distortion can affect both the virtual image of the display (the \emph{augmented view}), as well as the view of the real world as seen through the combining optics (the \emph{direct view}).  

For correcting the augmented view, Lee and Hua \cite{lee2013robust} propose a camera-based calibration method, that learns a corrective 2D distortion map in screen image space.  For correcting the direct view, Itoh and Klinker~\cite{itoh2015light} propose modeling the distortion as the shift in a bundle of 4D light rays (light field) passing through the optics, and then estimating a 4D-to-4D mapping between the original and distorted light fields.  Because it uses light fields, this method can handle viewpoint-dependent distortions.  Itoh and Klinker~\cite{itoh2015simultaneous} then extend this method to correct distortions of the augmented view.  Their evaluation with an OST HMD shows that removing both direct-view and augmented-view distortions provides overall registration accuracy comparable to 20/50 visual acuity.

Beyond the distortion estimation addressed by Itoh and Klinker~\cite{itoh2015light}, the same authors~\cite{itoh2016gaussian} further propose modeling the view-dependent color aberration (point spread function) of OST HMDs.  This method models the image blur as Gaussian functions integrated in the 4D-to-4D distortion mapping, and estimate it by measuring the impulse response of the display from different viewpoints.

\ParLabel{Summary}\label{sec:automatated_summary}
Clearly, automatic calibration methods are the future of OST HMDs.  In addition to freeing the human operator from having to manually perform calibration procedures, automatic methods could also operate in a closed-loop manner, continuously adjusting the calibration, and, therefore, correcting for small movements of the HMD on the user's head.  In addition, integrating eye trackers into an OST HMD allows many useful interaction techniques, such as gaze-based interaction, and also allows optimized rendering methods, such as foveated rendering.  However, as discussed, automatic calibration methods still face challenges, especially related to both eye models and display models.  

\section{Evaluation}\label{sec:eval}

It is, of course, important to evaluate calibration procedures.  However, especially compared to video see-through AR, evaluation in an OST HMD is particularly challenging, because, in the end, only the user can assess the locational realism of the result. This section summarizes existing evaluation methods.  Specifically, investigations have examined various data collection schemes, which have considered the presence or absence of postural sway, the effects of confirmation methods on dependent variables such as reprojection error, intrinsic and extrinsic parameter estimation, task completion time, and workload.  Table~\ref{tab:dc} summarizes these methods.  


In 2000, Genc \etal\cite{genc2000optical} evaluated a stereo calibration method briefly with two users, but used a video-see-through system.  Hence, these results are not easily transferable to optical see-through systems.

In 2001, McGarrity \etal\cite{mcgarrity2001evaluation, mcgarrity2001new} presented a method for providing registration accuracy feedback, where, using a stylus on a tablet, users indicate perceived positions of virtual objects.  They also propose using both 3D input points and their projected 2D point correspondences to adjust the calibration, but they do not provide an accuracy analysis of their approach.  Navab \etal\cite{navab2004line} later applied the idea to personal digital assistants (PDAs), and suggested using a point-and-shoot game to motivate users to complete calibration tasks.

 
In 2003, Tang \etal\cite{tang2003evaluation} compared 4 variants of SPAAM: SPAAM, DepthSPAAM (modified SPAAM to collect different depth values by moving the whole body relative to the 3D target), Stereo-SPAAM, and Stylus-Mark calibration (DepthSPAAM using a tracked stylus).  They focus on task completion time and geometric error (measured using the procedure described in McGarrity \etal\cite{mcgarrity2001evaluation, mcgarrity2001new}), and presented results for the decomposed principal point. They found that SPAAM resulted in the fastest task completion time but had the largest calibration error, while the Stylus-mark calibration had the lowest error.  However, the authors also note that ``none of the four procedures can achieve a reliable and accurate result for naive users''.

In 2008, Grubert \etal\cite{grubert2010comparative, grubert2008untersuchungen} compared a Multiple Point Active Alignment scheme (MPAAM) with SPAAM, and found that although the MPAAM calibration procedure significantly speeds up the data collection phase, it also results in larger calibration errors. 

In 2010, Axholt \etal\cite{axholt2010optical} used Monte-Carlo simulation to investigate the effects of human alignment noise on view parameter estimation.  Compared to a camera on a tripod, which can be perfectly still, a standing human will exhibit involuntarily postural sway, even if they attempt to stand perfectly still.  They found that the relatively large alignment noise induced by humans ($>$5px), compared to the lower alignment error typically reported in the computer vision literature for camera calibration (ca.~1px), primarily led to estimation variance in the extrinsic parameters along the user's line of sight ($z$ direction).  To mitigate this effect, they found that distributing the 3D correspondence points over a greater range of depths was more effective than simply adding additional correspondence points. 

Subsequently, Axholt \etal\cite{axholt2011parameter} investigated the effects of 3D point distribution patterns.  They compared \emph{static $z$} (a single $z$ depth distance), \emph{sequentially increasing $z$} (resulting in an upward curved trapezoidal shape), and \emph{magic square} (systematic variance in $x, y$ such that $z$ depth changes are maximized) acquisition patterns. The authors found that the magic square pattern resulted in the least parameter variance.  They also found that orientation and lateral principal point offset are not primarily affected by the correspondence point distributions, but depend on the number of correspondence points. 

In his 2011 dissertation, Axholt~\cite{axholt2011pinhole} further summarized the main findings of several studies on the influence of human alignment noise on OST HMD calibration.  He had several main findings: First, for standing users completing a calibration task based on visual alignment, postural stability gives a translational head-aiming precision of 16 mm, which improves to 11 mm after 12--15 seconds, and can be modeled with a Weibull distribution.  Second, for standing users, head aiming precision is 0.21$^\circ$ straight ahead and 0.26$^\circ$ in directions $\geq$ 30$^\circ$ azimuth, but the precision can be improved by considering postural sway and compensatory head rotation together, resulting in a precision of 0.01$^\circ$.  For seated users, the precisions are 0.09$^\circ$ in directions $\leq$ 15$^\circ$ azimuth, and can be approximated with a circular distribution.  Third, pinhole camera parameter estimation variance increases linearly as a function of both alignment noise and diminishing correspondence point depth distribution.  It decreases optimally if 25 or more correspondence points are used. It also decreases for all parameters with increasing correspondence point depth distribution variance, except for rotation, which primarily depends on the number of correspondence points.  Finally, for seated subjects using SPAAM and a pinhole camera model, the eye point estimation accuracy is only 5cm on average, and depends on the camera matrix decomposition method used (none; closed form solution as described by Faugeras~\cite{faugeras1993three} (p.~52) as well as Trucco and Verri~\cite{trucco1998introductory} (p.~134); RQ decomposition using Givens rotations (Hartley and Zisserman~\cite{hartley2003multiple}, p.~579)). 


In 2011, Maier \etal\cite{maier2011empiric} investigated how different confirmation methods affect calibration quality.  They compared keyboard, hand held, voice, and waiting methods, and found that the waiting method was significantly more accurate than the other methods.  They also found that averaging the data collection over time improved the accuracy of all methods.  However, their experiment used a video see-through HMD, and so the results could differ for an OST HMD. 

In 2014, Moser \etal\cite{moser2014baseline} conducted an experiment to generate baseline accuracy and precision values for OST HMD calibration, without human postural sway error.  To this end, the authors mounted a camera inside an OST HMD, used SPAAM, and compared the same three depth distributions as Axholt \etal\cite{axholt2011parameter}: static $z$, sequentially increasing $z$, and magic square.   Replicating Axholt \etal\cite{axholt2011parameter}, they found that the magic square pattern, which yields greater depth variance for the same number of correspondence points, produces the most accurate and precise results.

In 2014, Itoh and Klinker~\cite{itoh2014interaction, itoh2014performance} analyzed the error sensitivity of SPAAM, DSPAAM (a degraded version of SPAAM where actual display use is simulated by removing and then replacing the display on the head), and the recycled / full INDICA method.  For each calibration method, they simulated how errors in calibration parameters propagate to the final calibration result.  Their analysis shows that, for both INDICA methods, the display orientation with respect to the HMD coordinate system has the largest impact on the reprojection error.   In addition, they confirmed that SPAAM tends to provide erroneous eye positions along the $z$ viewing direction.  Note that the DSPAAM method simulates a common scenario in consumer applications, where non-expert users rely on a factory calibration or only perform the calibration once.  This use pattern creates several errors; for example, every time the user puts on the HMD, the alignment between the display screens and the eyes varies slightly.  

In 2015, Moser \etal\cite{moser2015subjective} compared SPAAM, DSPAAM, and recycled INDICA, using both objective and subjective evaluations of two tasks: (1) indicating the location of a virtual pillar, 15.5cm high, seen against a 4x4 grid of physical pillars, and (2) indicating the location of a 2x2x2cm virtual cube on a physical 20x20x20cm grid.  They found no significant differences between SPAAM and DSPAAM, and no difference between any of the methods in the left / right $x$ axis.  They found that recycled INDICA resulted in the best accuracy in the depth $z$ and up / down $y$ axes.  Also, recycled INDICA resulted in the highest subjective user preference, likely because it requires minimal user effort and took the least time to perform.  Finally, the authors note that there was a substantial disagreement between subjective and objective measures, because depth errors are less easily perceived than left / right or up / down errors. 


In 2016, Moser \etal\cite{moser2016evaluation} evaluated the feasibility of performing SPAAM calibration using a Leap Motion controller\footnote{www.leapmotion.com/ - last accessed September 2nd, 2017.} as a 3D input tool, similar to Tang \etal's~\cite{tang2003evaluation} use of a 3D tracked stylus.  Moser \etal\cite{moser2016evaluation} compared four tracked objects: the user's finger matched to a virtual cross, box, and finger-shaped reticle, and a wooden stylus matched to a virtual cross.  SPAAM calibrations were performed in both monoscopic and stereo conditions.  A single expert user performed 20 calibrations for each of the 8 object-by-stereopsis conditions, where calibration required matching 25 calibration points.  For dependent measures they evaluated both reprojection error and eye location estimates, and for the stereo calibrations they additionally evaluated binocular $x, y, z$ disparities.  Note that the $x$ binocular disparity measures the expert user's inter-pupillary distance, and because this number is independently measurable, the inter-pupillary distance is an excellent metric for evaluating the accuracy of a stereo calibration procedure.  For all dependent measures, Moser \etal\cite{moser2016evaluation} found that stylus calibrations were were much more accurate than all of the finger methods.  They attributed this finding to the Leap Motion controller's relatively low accuracy finger tracking. 

Also in 2016, Jun and Kim \cite{jun2016calibration} evaluated their calibration method for stereo calibration using a depth-camera against stereo-SPAAM \cite{genc2000optical}. They found their model to perform better (in terms of positional error) with a fewer number of point correspondences.

Zhang \etal \cite{zhang2017ride} compared their RIDE method with standard SPAAM within a grid of 5 x 5 = 25 sampling points and found their method to result in a lower reprojection error (3.36 pixel for RIDE vs. 5.29 pixel for SPAAM in a 800x600 px display. However, they only used a single user who performed three repetions per method.

Additionally, Qian \etal\cite{qian2016stereo} proposed additional constraints for Stereo SPAAM calibration utilizing known properties and physical presumptions about the physical structures of the user's eyes. These constraints include the assumption of identical pixel density in both the x and y axis along the screen for each eye, no skew in the perceived image, identical viewing direction of both eyes perpendicular to the imaging plane, and that the interpupillary distance can be measured and known. Reprojection error is used for the comparative metric, and their results show that the inclusion of these additional constraints show promise in reducing calibration errors for binocular systems that are able to conform to the necessary restrictions. A larger study is still needed, however. 

A second work by Qian \etal\cite{qian2016fixed} examines the use of physical head constraint during SPAAM, akin to bore-sighting, to reduce the number of free parameters needed for estimation. In this study, the user's head is fixed to translation and rotational error with alignments being controlled by a mouse pointer on a physical display screen. As with the previous study, this work also shows a reduction in calibration error, in terms of reprojection error, with the trade-off of head restriction.

Azimi \etal\cite{azimi2017anisotropic} in 2017, proposed a new metric for both quantifying and improving calibration accuracy with regards to reprojection error. They propose the use of Mahalonobis distance instead of the traditional Euclidean distance in measuring reprojection offset of SPAAM results. Their reasoning lies in the anisotropic nature of the user alignment data during the calibration. Through the application of Mahalonobis distance measures, a post calibration iterative error reduction process is applied to identify and reduce the number of user alignment outliers. The results of their study show statistically significant improvements in both final reprojection error and reduction in required user alignments using this new metric.


\begin{table*}[t]
    \caption{Overview of evaluation methods. \textbf{Methods Evaluated}: \textbf{DC}: Data collection approach (\textbf{O}: objective measures, \textbf{SQN}: Subjective quantitative, \textbf{SQL}: subjective qualitative), \textbf{TCT} (task completion time), \textbf{WL} (workload), \textbf{GE} (geometric errors: reprojection or viewing angle), \textbf{PRE} (parameter reconstruction error from projection matrix, e.g., eye location).
    \vspace{-3.5mm}
    }
    \label{tab:dc}
\begin{center}
    \begin{tabular}{p{0.17\textwidth} p{0.35\textwidth} p{0.05\textwidth} p{0.025\textwidth} p{0.025\textwidth} p{0.025\textwidth} p{0.025\textwidth} p{0.1\textwidth}}
    
    \toprule
    \multicolumn{2}{c}{} & \multicolumn{6}{c}{\bf{Measures (dependent variables)}} \\ 
    \cmidrule(l){3-8}

    \textbf{Work} &\textbf{Methods Evaluated} & \textbf{DC} & \textbf{TCT} & \textbf{WL} & \textbf{GE} & \textbf{PRE} & \textbf{Other}\\ \midrule
    
    McGarrity \etal\cite{mcgarrity2001evaluation,mcgarrity2001new} & SPAAM & SQN & x & x & \checkmark & x & x\\ \midrule
    
    Tang \etal\cite{tang2003evaluation} & \raggedright SPAAM, DepthSPAAM head pointing, DepthSPAAM stylus pointing, Stereo-SPAAM & SQN & \checkmark & x & \checkmark & \checkmark & x \\ \midrule
    
    Grubert \etal\cite{grubert2010comparative, grubert2008untersuchungen} & DepthSPAAM, MPAAM & SQN & \checkmark & x & \checkmark & x & x \\ \midrule
    
    Axholt \etal\cite{axholt2010optical} & DLT (simulated point correspondences) & O & x & x & x &  \checkmark & x\\ \midrule
    
    Axholt \etal\cite{axholt2011parameter} & \raggedright SPAAM, DepthSPAAM Sequential, DepthSPAAM Magic Square & O & x & x & x & \checkmark & condition number \\ \midrule
    
    Maier \etal\cite{maier2011empiric} & \raggedright SPAAM with 4 confirmation methods: keyboard button, handheld button, voice input, waiting & O & x & x & \checkmark  & x & x \\ \midrule
    
    Moser \etal\cite{moser2014baseline} & \raggedright SPAAM, DepthSPAAM Sequential, DepthSPAAM Magic Square & O & x & x & x & \checkmark & x \\ \midrule
    
    Moser \etal\cite{moser2015subjective} & SPAAM, Degraded SPAAM, Recycled INDICA & SQN, SQL & x & x & \checkmark & \checkmark & x\\ \midrule
    
    Itoh and Klinker\cite{itoh2014performance} & \raggedright SPAAM, Degraded SPAAM, Recycled/Full INDICA & O & x & x & \checkmark & x & perturbation sensitivity\\ \midrule
    
    Moser and Swan\cite{moser2016evaluation} & \raggedright SPAAM with head pointing, finger pointing, stylus pointing (mono + stereo) & SQN & x & x & \checkmark & \checkmark & x\\ \midrule
    
    Qian \etal\cite{qian2016fixed} & \raggedright SPAAM with fixed head, mouse pointer alignment & SQN & x & x & \checkmark & x & x \\ \midrule 
    
    Qian \etal\cite{qian2016stereo} & \raggedright Stereo SPAAM, head pointing & SQN & x & x & \checkmark & x & x \\ \midrule
    
    Zhang \etal \cite{zhang2017ride} & \raggedright RIDE, SPAAM & SQN & x & x & \checkmark & x & x \\ \midrule
    
    Azimi \etal\cite{azimi2017anisotropic} & \raggedright SPAAM with head pointing and Mahalonobis distance error correction & O & x & x & \checkmark & x & Mahalonobis distance \\ \midrule
    
    Jun and Kim \cite{jun2016calibration} & \raggedright Proprietary methods (full, simplified) vs. Stereo SPAAM & O, SQL & x & x & x & x & 3D positional error, qualitative image results \\ \bottomrule

    

    \end{tabular}
    \end{center}

\end{table*}

\section{Opportunities for Future Research}
Future directions for research in optical see-through calibration methods can be identified in the areas of error metrics, display models, eye trackers and methods that go beyond spatial calibration.  

\ParLabel{Improving Error Metrics} 
Almost all of the reviewed papers report calibration reprojection errors in pixels (px).  However, because HMDs have different resolutions, and because the distance between the user's eyes and the virtual screen plane varies by both HMD and user, reporting errors in pixels makes it difficult to meaningfully compare different methods.  To address this, we propose expressing calibration errors in degrees of visual angle.  In addition to being a more comparable unit, degrees of visual angle is the standard unit for many results in the vision science community.  In addition, most evaluation methods only report performance metrics.  However, manual calibration can be a strenuous task, and we therefore advise using subjective workload measurements (e.g., NASA TLX \cite{hart1988development}), as well as additional measures focusing specifically on eye strain (c.f., oculumotor component of the Simulator Sickness Questionnaire \cite{kennedy1993simulator}). In addition, objective stress measures could be employed. In contrast to evaluating effects of HMD usage on user comfort \cite{grubert2010extended}, there have been no evaluations on long term effects of HMD slippage in real-world scenarios. These investigations could help to understand the practical boundaries that automated calibration methods would need to account for.  Finally, as discussed in Section~\ref{sec:eval}, when possible, we recommend reporting a calibration technique's measurement of interpupillary distance, because this metric is user-centered, varies between users, is an important graphical parameter, and for each user can be independently measured with a high degree of accuracy.  

\ParLabel{Advanced Display Models} 
The vast majority of the methods reviewed here model the OST HMD as an off-axis pinhole camera, introduced in Section 2.  Although this assumption is plausible for OST HMDs that use an optical combiner and virtual screen plane, additional accuracy is likely possible by using more complex graphical and eye models (Axholt~\cite{axholt2011pinhole}, Jones \etal\cite{jones2016schematic}).  In addition, other display types call for extending the current display model.  For example, focus-tunable OST HMDs can move the screen to an arbitrary focus depth (Liu \etal\cite{liu2010novel}, Hu and Hong~\cite{hu2014design}, \cite{hu2014high}, and Dunn \etal\cite{dunn2017wide}).  In such displays, the display model must always represent the current screen focus depth.  Light field displays create virtual images with variable accommodation (Maimone and Fuchs~\cite{maimone2013computational}), but to update the virtual image, these displays also need to know the user's eye position.  

Other recent HMD systems use phase-only liquid crystal on silicon (P-LCOS) displays. P-LCOS displays are 
essentially a programmable lens mirror that can change the refraction index of each pixel independently. 
The surface focal displays use P-LCOS to create a dynamic depth field in VR HMDs (Matsuda \etal\cite{matsuda2017focal}). A true holographic OST HMDs was developed by using P-LCOS displays
(Maimone \etal\cite{maimone2017holographic}). These are promising approaches once we make P-LCOS small
enough to be integrated in HMDs.
%



\ParLabel{Integrating Eye Trackers} 
As discussed in Section~\ref{sec:auto}, many current methods seek to automate the calibration process, but these methods will require eye trackers seamlessly integrated into the OST HMD.  Therefore, integrating an eye tracker with an acceptable form factor is an important issue.  Hua~\etal~\cite{hua2013high} prototyped an OST HMD with an IR eye tracker that is integrated in the optics, and partially shares the optical path with the display.

Even after integrating an eye tracker (eye camera), for automatic calibration, one still needs to calibrate its pose in the display coordinate system.  The pose estimation, however, can be challenging. If the coordinate system is defined on a scene camera, one needs to calibrate the pose between the scene camera and the eye camera, where the two cameras are looking in opposite directions, at the world and at the eye, with extremely different focal lengths, in the range of meters and centimeters (Itoh and Klinker~\cite{itoh2014interaction}).  Some researchers report estimating the pose via visual marker tracking, with a custom multi-marker jig (Itoh and Klinker~\cite{itoh2014interaction,itoh2014performance}).

In an outside-in tracking setup, where the display coordinate system is defined on a set of optical markers attached on the display, the calibration procedure could even be more complex.  In such systems, the external outside-in tracking system may not be able to track the jig, and, therefore, a marker jig may not work.  Instead, a hand-eye calibration must be applied between the marker set and the eye camera (Horaud and Dornaika~\cite{horaud1995hand}).

Furthermore, the eye tracker pose might change during the use of the OST HMD because the user may touch the eye camera or the camera needs to be re-oriented to capture the eyes properly. As a result, the eye tracker could require frequent re-calibration. Plopski \etal~\cite{plopski2016automated} propose to automatically calibrate the pose via the corneal reflection of LED arrays attached on both the eye tracker and the display frame. The integrated design from Hua~\etal~\cite{hua2013high} mentioned above may also be another hardware solution. Since the eye camera image is frontal to the eye, the camera inside the HMD frame could be fixed and only calibrated once, possibly at the manufacturer side.


\ParLabel{Beyond Spatial Calibration} 
Throughout the paper, we looked at various existing calibration methods that aim at improving the alignment accuracy of AR images against the physical world in the user's field of view. A question, which arises, is, how accurate do calibrating methods need to become. 
Logically thinking, the maximum accuracy would end up to the \textit{retinal-precise} accuracy where an OST HMD can align each pixel of a displayed AR image to desired retina cells. In other words, the display can stimulate arbitrary retinal cells selectively with desired light stimulation.
Such accuracy might be overkill for most AR applications. However, if such calibration accuracy could be achieved, OST HMDs may go beyond the realm of AR displays---they could be devices that can arbitrarily manipulate human vision. A potential application of such direction is vision augmentation, where AR displays enhance human vision by retinal-precise image processing. There already exist a few applications that demonstrate such vision augmentation concepts with OST HMDs (Itoh and Klinker~\cite{itoh2015vision}, Hwang and Peli~\cite{hwang2014augmented}).


\section{Conclusion}

This paper surveyed the field of calibration methods for OST HMDs. Specifically, it reviewed approaches accessible until September 2017. It provided insights into the fundamentals of calibration techniques, manual and automatic  calibration approaches, as well as evaluation methods. These calibration methods are focused achieving on locational realism, that is, the correct spatial alignment of virtual content in a physical environment. Besides locational realism, further consistency domains in addition to the spatial domain, including color calibration (Itoh \etal~\cite{itoh2015semi}, Langlotz \etal~\cite{langlotz2016real}) and latency (Lincoln \etal~\cite{lincoln2016motion}), should be considered to achieve a as high degree of perceived realism as possible.

\section*{Acknowledgments}


This material is based upon work supported by the National Science Foundation under awards IIS-1018413 and IIS-1320909, to J. E. Swan II, and fellowships provided by the NASA Mississippi Space Grant Consortium and Japan Society for the Promotion of Science (JSPS) through the East Asia and Pacific Summer Institutes Fellowship, award IIA-141477, to K. Moser, and JSPS KAKENHI Grant Numbers 16H07169, 17H04692, and 17K19985, and European Union’s 7th Framework Programmes for Research and Technological Development under PITN-GA-2012- 316919 EDUSAFE, to Y. Itoh.

\ifCLASSOPTIONcaptionsoff
  \newpage
\fi



\bibliographystyle{IEEEtranS} 
\bibliography{IEEEabrv,./Bib/ostbib}

\begin{IEEEbiography}[{\includegraphics[width=1in,height=1.25in,clip,keepaspectratio]{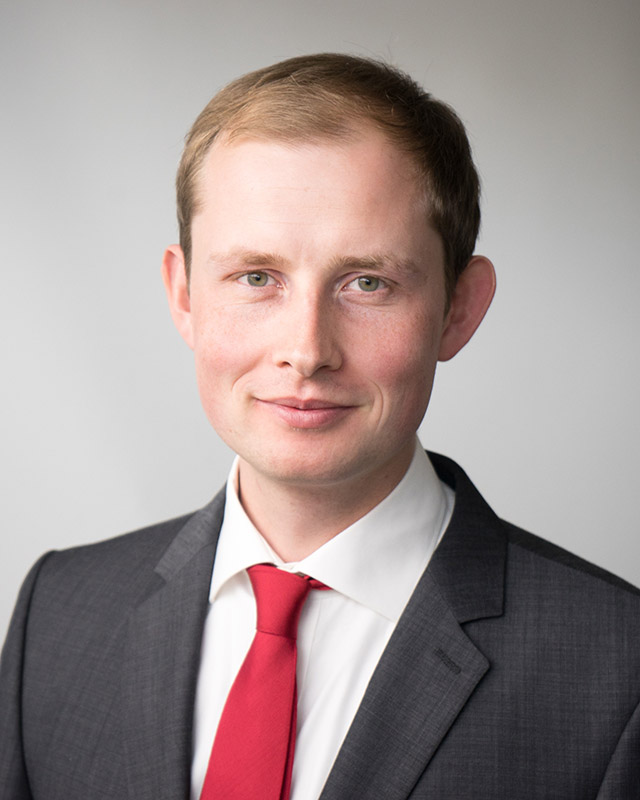}}]{Jens Grubert}
(Member IEEE) is Associate Professor for Human-Computer Interaction in the Internet of Things and lab director of the mixedrealitylab, a laboratory for Augmented and Virtual Reality, at Coburg University, Germany. Previously, he held positions at University of Passau, Graz University of Technology, University of Applied Sciences Salzburg, Austria and Fraunhofer Institute for Factory Operation and Automation IFF, Germany. He received his Dr. techn  (2015) with highest distinction at Graz University of Technology, his Dipl.-Ing. (2009) and Bakkalaureus (2008) with highest distinction at Otto-von-Guericke University Magdeburg, Germany. He is author of more than 50 peer reviewed publications and patents and published a book about AR development for Android. His current research interests include interaction with multimodal augmented and virtual reality, body proximate display ecologies, around-device interaction, multi-display environments and cross-media interaction. 
\end{IEEEbiography}

\begin{IEEEbiography}[{\includegraphics[width=1in,height=1.25in,clip,keepaspectratio]{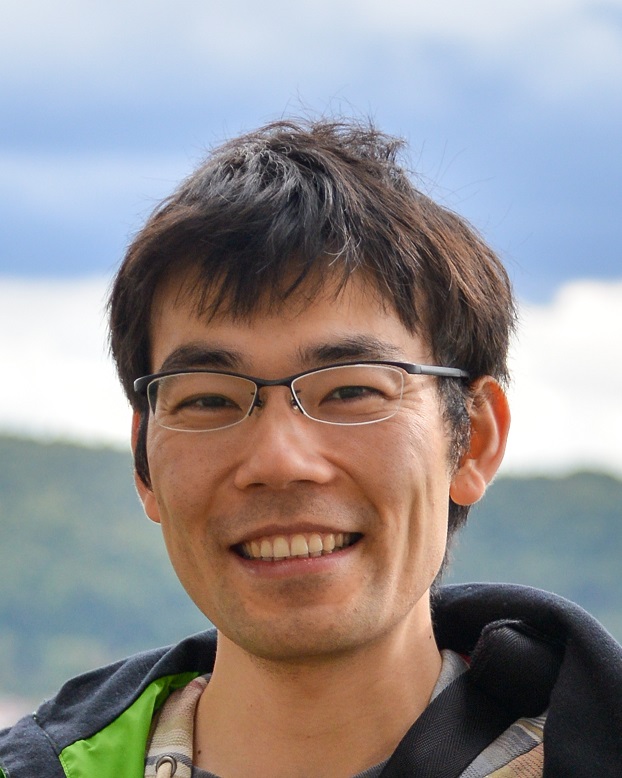}}]{Yuta Itoh}
(Member IEEE) is Yuta Itoh is a project assistant professor at Keio University, Japan. 
He received his Dr. rer. nat. from Technical University of Munich (2016).
He holds B.Eng. (2008) and M.Eng. (2011) degrees in computer science
from Tokyo Tech, where he studied machine learning. He spent two years as a researcher at Multimedia Lab. in Toshiba Corp.
(2011-2013). His research topic is in vision augmentation which aims to support and enhance human vision via
augmented reality, especially with optical see-through near-eye displays.
\end{IEEEbiography}

\begin{IEEEbiography}[{\includegraphics[width=1in,height=1.25in,clip,keepaspectratio]{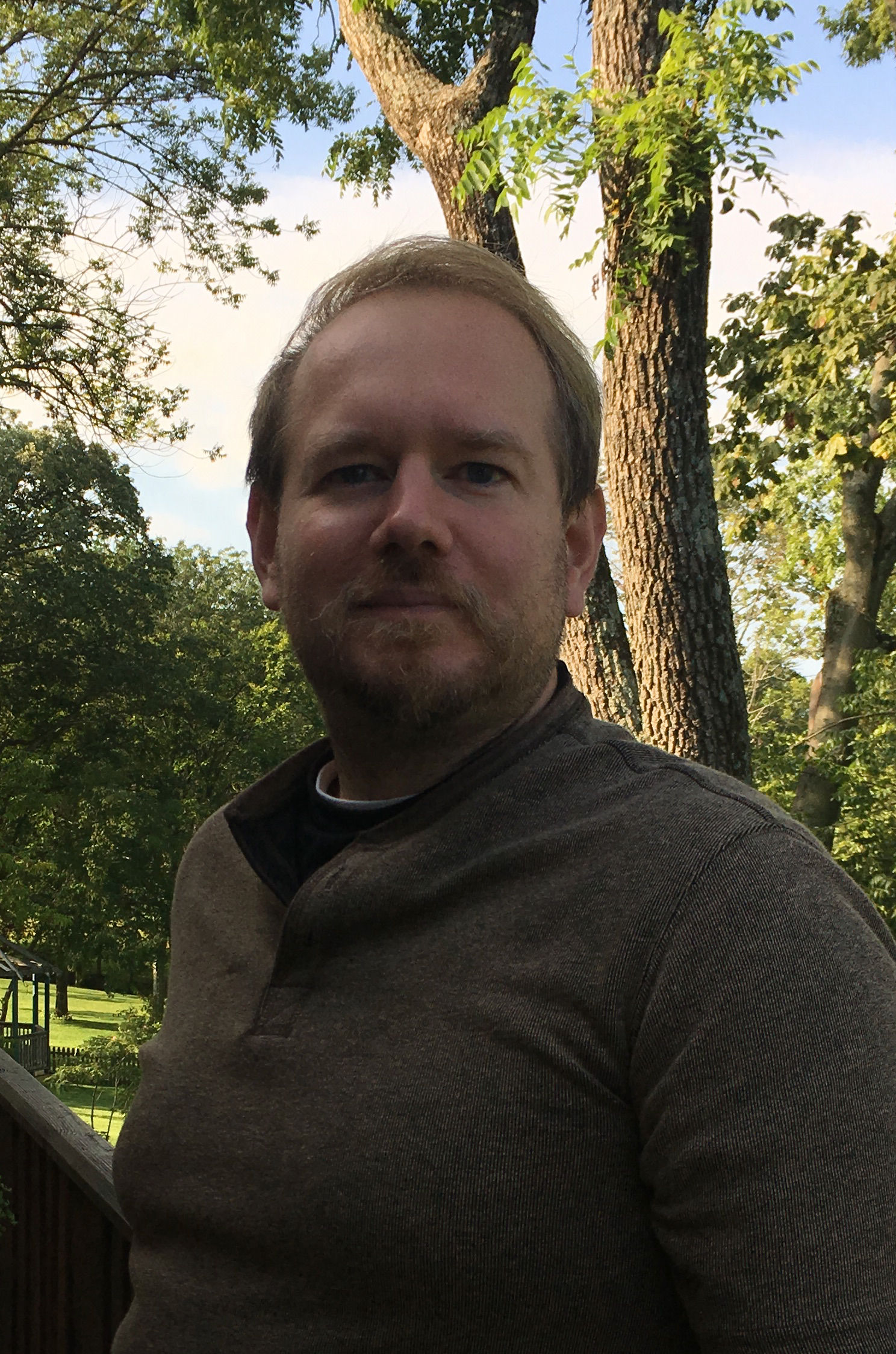}}]{Kenneth Moser}
is a Research and Development Engineer at Marxent Labs LLC, in Dayton Ohio, USA. He received his PhD. in Computer Science from Mississippi State University (2016) and holds a M.S. (2011) in Computer Science and B.S. (2005) in Mechanical Engineering from Mississippi State University. His doctoral studies included two study abroad projects at the Interactive Media Design Lab, at the Nara Institute of Science and Technology in Nara, Japan as well as a summer internship project at the Naval Research Laboratory in Washington, D.C. His current research focus is creating consumer AR and VR experiences for e-commerce.
\end{IEEEbiography}

\begin{IEEEbiography}[{\includegraphics[width=1in,height=1.25in,clip,keepaspectratio]{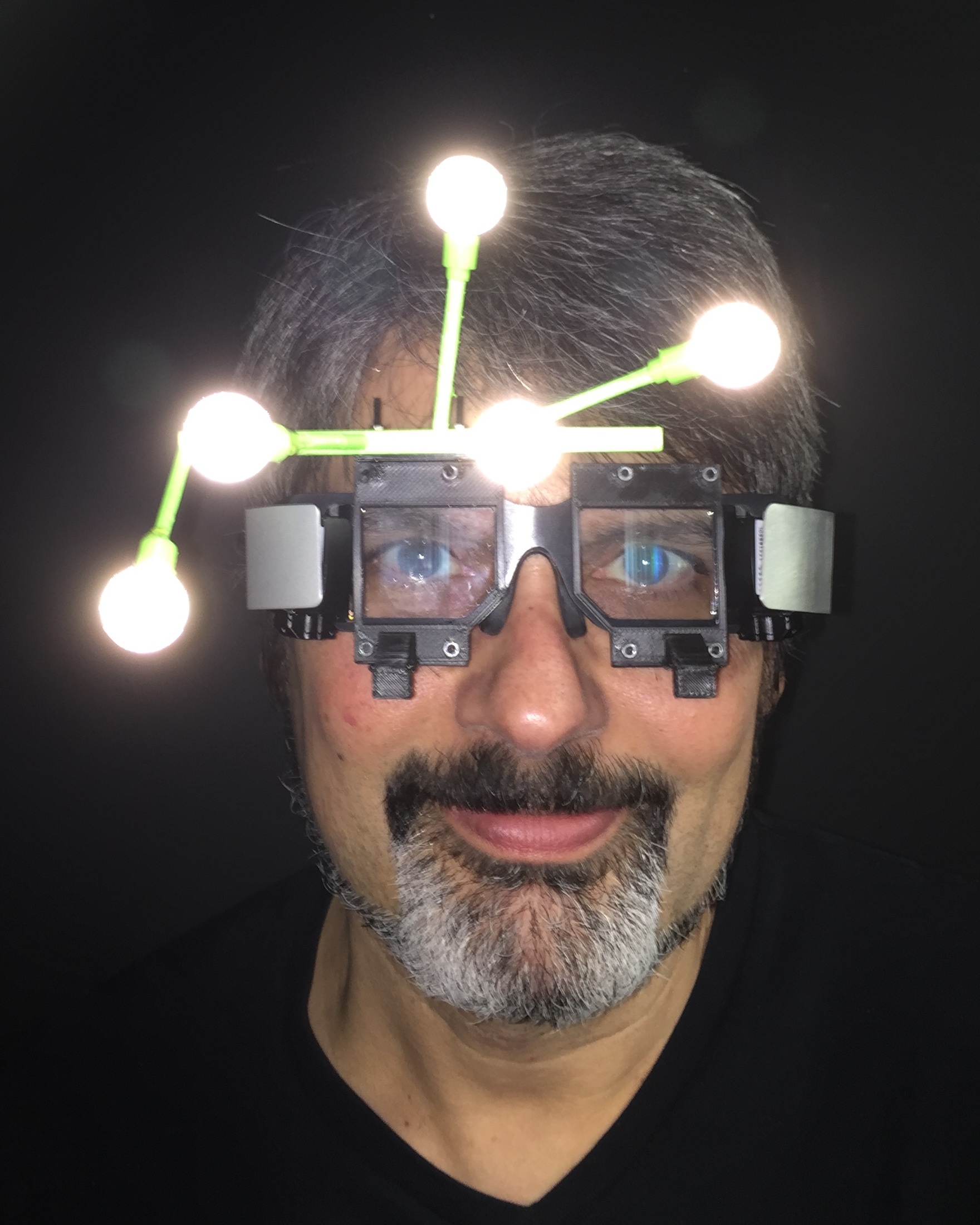}}]{J. Edward Swan II}
(Senior Member IEEE) is a Professor and Interim Department Head of Computer Science and Engineering, and Faculty at the Center for Advanced Vehicular Systems, at Mississippi State University.  He holds a B.S. (1989) degree in computer science from Auburn University and M.S. (1992) and Ph.D. (1997) degrees in computer science from Ohio State University, where he studied computer graphics and human-computer interaction.  Before joining Mississippi State University in 2004, Dr. Swan spent seven years as a scientist at the Naval Research Laboratory in Washington, D.C.  Dr. Swan’s research and scholarship has encompassed augmented and virtual reality, perception, human-computer interaction, human factors, empirical methods, computer graphics, and visualization.  Currently, Dr. Swan is studying perception and calibration in augmented and virtual reality, including depth and layout perception and depth presentation methods.  His research has been funded by the National Science Foundation, the Department of Defense, the National Aeronautics and Space Administration, the Naval Research Laboratory, and the Office of Naval Research.  Dr. Swan is a member of ACM, IEEE, the IEEE Computer Society, and ASEE.
\end{IEEEbiography}

\newpage






\end{document}